\documentclass[12pt]{article}
\usepackage{amsmath,amsfonts}
\usepackage[american]{babel}
\usepackage{mathrsfs}

\def\cstok#1{\leavevmode\thinspace\hbox{\vrule\vtop{\vbox{\hrule\kern1pt
\hbox{\vphantom{\tt/}\thinspace{\tt#1}\thinspace}}
\kern1pt\hrule}\vrule}\thinspace}

\title{Spectral methods in quantum field theory and quantum cosmology}

\author{Giampiero Esposito$^{1}$, Guglielmo Fucci$^{2}$, \\
Alexander Yu. Kamenshchik$^{3,4,5}$
and Klaus Kirsten$^{2}$,\\
${ }^{1}$INFN Sezione di Napoli,\\
Complesso Universitario di Monte S. Angelo,\\
Via Cintia, Edificio 6, 80126 Napoli, Italy,\\
${ }^{2}$Department of Mathematics, Baylor University,\\
Waco, TX 76798 USA,\\
${ }^{3}$Dipartimento di Fisica,
Via Irnerio 46, 40126 Bologna, Italy,\\
${ }^{4}$INFN, Sezione di Bologna,\\
Via Irnerio 46, 40126 Bologna, Italy,\\
${ }^{5}$L.D. Landau Institute for Theoretical Physics, \\
Russian Academy of Sciences,
Kosygin str. 2, 119334 Moscow, Russia}

\date{\today}

\begin{document}

\maketitle

\begin{abstract}
We review the application of the spectral zeta-function to the
1-loop properties of quantum field theories on manifolds with
boundary, with emphasis on Euclidean quantum gravity and quantum
cosmology. As was shown in the literature some time ago, the only
boundary conditions that are completely invariant under infinitesimal
diffeomorphisms on metric perturbations suffer from a drawback, i.e.
lack of strong ellipticity of the resulting boundary-value problem.
Nevertheless, at least on the Euclidean 4-ball background, it remains
possible to evaluate the $\zeta(0)$ value, which describes in this
case a universe which, in the limit of small 3-geometry, has vanishing
probability of approaching the cosmological singularity. An assessment
of this result is here performed, discussing its physical and
mathematical implications.
\end{abstract}

\section{Introduction}

In the Euclidean functional-integral approach to quantum
gravity, one deals with amplitudes written formally
as functional integrals over all Riemannian 4-geometries
matching the boundary data on (compact) Riemannian
3-geometries $\Bigr(\Sigma_{1},h_{1}\Bigr)$ and
$\Bigr(\Sigma_{2},h_{2}\Bigr)$ \cite{M1957}.
To take into account the
gauge freedom of the theory, the functional-integral measure
also includes suitable ghost fields, described geometrically
by a 1-form, hereafter denoted by $\varphi=\varphi_{\mu}
dx^{\mu}$, subject to boundary conditions at
$\Bigr(\Sigma_{1},h_{1}\Bigr)$ and $\Bigr(\Sigma_{2},h_{2}\Bigr)$.
Although a rigorous definition of the Feynman sum over all
Riemannian 4-geometries with their topologies does not
yet exist, the choice of boundary conditions still plays a key
role to obtain an elliptic boundary-value
problem, which may be applied to the semiclassical analysis of
the quantum theory.

In quantum cosmology, it was proposed in \cite{HH1983} and \cite{H1984}
that no boundary conditions should be imposed at the 3-geometry
$\Bigr(\Sigma_{1},h_{1}\Bigr)$, since this might shrink to a
point in the case of the quantum state of the universe. One
would then have to impose suitable boundary conditions only
at $\Bigr(\Sigma_{2},h_{2}\Bigr)$, by describing the quantum
state of the universe in terms of an Euclidean functional integral
over all {\it compact} Riemannian 4-geometries matching
the boundary data at $\Bigr(\Sigma_{2},h_{2}\Bigr)$. Although
this approach to quantum cosmology still involves a number
of formal definitions, the semiclassical evaluation of the
corresponding wave function may be put on solid ground.
The 1-loop analysis
is related to mathematical and physical subjects such
as cobordism theory (i.e., under which conditions a compact
manifold is the boundary of another compact manifold),
the geometry of compact Riemannian
4-manifolds, the asymptotic heat kernel, the 1-loop effective
action, and the use of mixed boundary conditions in
quantum field theory (see below).

In particular, over the last decades many efforts have been
produced to evaluate 1-loop
quantum amplitudes for gauge fields and the gravitational
field in the presence of boundaries, either by using the
space-time covariant Schwinger-DeWitt method or the mode-by-mode
analysis which relies on zeta-function regularization.
The main motivations were the
need to understand the relation between different approaches to
quantum field theories in the
presence of boundaries and the quantization of closed cosmologies.
Indeed, boundaries play an important role in the Feynman
approach to quantum gravity as we just said, in choosing
Becchi-Rouet-Stora-Tyutin- (BRST-) covariant
and  gauge-invariant boundary conditions
for quantum cosmology and in studying different quantization
and regularization techniques in field theory.
In particular, for the latter problem,
discrepancies were found in the semiclassical evaluation of
quantum amplitudes by using space-time covariant methods, where
the scaling factor of 1-loop quantum amplitudes
coincides with the Schwinger-DeWitt
$A_{2}$ coefficient in the heat-kernel expansion, or instead a
mode-by-mode analysis, for which the resulting equations obeyed
by the eigenvalues are studied through
zeta-function methods \cite{DC1976,H1977}.

If one reduces a field theory with first-class constraints
\cite{HT1992} to its
physical degrees of freedom before quantization,
one of the main problems is whether
the resulting quantum theory is equivalent to the
theories relying on the Faddeev-Popov gauge-averaging method or
on the extended-phase-space Hamiltonian functional integral of Batalin,
Fradkin, and Vilkovisky, where one takes into account ghost and
gauge modes. We will see that, in a mode-by-mode evaluation
of the covariant functional integral including gauge-averaging
and ghost terms, after doing a 3+1 split and a Hodge-like
decomposition of the components of metric and ghost
perturbations, there are no exact cancellations between
contributions of gauge and ghost modes,
when linear covariant gauges are used.
This lack of cancellation turns out to be essential to
achieve agreement between different techniques.

In \cite{S1985}, the $\zeta(0)$ calculation was performed
for gravitons by restricting the functional-integral measure
to transverse-traceless perturbations
in the case of flat Euclidean 4-space bounded by a 3-sphere.
In \cite{BKK1992,BKKM1992,KM1992} this result was
generalized to the part of the Riemannian de Sitter
4-sphere bounded by a 3-sphere. Both results did
not coincide with those obtained by a space-time covariant
method \cite{MP1990}.
Hence the natural hypothesis arises that the
possible non-cancellation of the contributions of gauge
and ghost modes can be the cause of the discrepancy.
In the work presented in \cite{EKMP1994,EKMP95a}
such a suggestion was checked
for the electromagnetic field on different manifolds and in
different gauges.

In \cite{BG1990} the asymptotic heat kernel for second-order
elliptic operators was obtained in the case
of pure and mixed boundary conditions in real Riemannian
4-manifolds, and in \cite{V1995} this analysis was
improved. In the light of these results,
the conformal anomalies on Einstein spaces with boundaries
were recalculated in \cite{MP1994}.

In \cite{EKMP95b} the linearized gravitational field was studied
in the geometric framework of \cite{S1985}
(i.e. flat Euclidean 4-space bounded by a 3-sphere),
and the resulting $\zeta(0)$ value was compared
with the space-time covariant calculation of the same Faddeev--Popov
amplitudes, by using the corrected geometric formulae
for the asymptotic heat kernel in the case of mixed
boundary conditions \cite{MP1994}.

However, in the case of mixed boundary conditions involving
tangential derivatives of metric perturbations, no geometric
formulae for the asymptotic heat kernel are available as yet,
and one has to resort, to the best of our knowledge, to analytic
techniques along the lines of the work in
\cite{EFKK05a,EFKK05b}. This is what
our review is mainly devoted to. For this purpose, section 2 derives
the integral representation of the spectral zeta-functions, as
obtained in \cite{BKK1992,bord96-37-895,BGKE1996}. 
Sections 3 and 4 are devoted to
$\zeta(0)$ values for scalar and gauge fields, respectively.
Detailed calculations for the gravitational field begin in section 5
and continue until section 7. The strong ellipticity issue is
studied in section 8, with examples. Concluding remarks and
open problems are presented in section 9.

\section{Integral representation of the spectral zeta-function}

A convenient method for the calculation of the spectral zeta function
for the case when the spectrum is not known explicitly,
but only the structure of the basis functions
of the corresponding differential operator are known,
was proposed in \cite{BKK1992}.
Here we sketch the basic ideas and formulae of this method; for a 
related approach see Dowker \cite{dowk96-13-585,dowk96-366-89}.

Let us consider the second-order operator $F$, which represents the
second functional derivative of the Euclidean action of
the model under consideration with respect to the field variables.
It is convenient to single out the mass term $m^2$ from the operator $F$.
As a manifold we consider the part of the closed Euclidean
de Sitter space (``Euclidean ball'').
Then suppose that we have a full set of basis
functions  $u^k_{A}(\tau|m^2)$ of this massive operator, i.e.
\begin{equation}
\left[F_{ik}\left(\frac{d}{d\tau}\right)+m^2a_{ik}\right] 
u_{A}^{k}(\tau |m^{2})= 0.
\label{basis}
\end{equation}
Here $\tau$ is the Euclidean time parameter, lower case Latin
indices $i$ enumerate the modes of the field variables
while capital Latin indices enumerate the basis functions,
$a_{ik}$ are the coefficients of the second-order derivatives
with respect to the time parameter.

The only condition which these basis functions should satisfy is
regularity in the Euclidean ball $0 \leq \tau \leq \tau_+$. Then the
eigenvalues $\lambda $ of the operator $F(d/d\tau) +m^2a$ with
homogeneous Dirichlet boundary conditions satisfy the equation
\begin{equation}
u_{A}^i(\tau_+|m^2-\lambda) = 0.
\label{eigen}
\end{equation}
For other types of boundary conditions the basis functions $u$
in  Eq. (\ref{eigen}) should be substituted by the corresponding
combination of basis functions and their derivatives.
For the case of Dirichlet boundary conditions, which we consider
in more detail in this section,
the equations defining all the eigenvalues
can be collected in one equation,
\begin{equation}
\det u_A^i(\tau_+|m^2-\lambda) = 0,
\label{det}
\end{equation}
in which the determinant is taken with respect to the indices $i$
and $A$ of the square matrix $u_A^i$. Then using the Cauchy formula
and the well-known relation between $\det$ and ${\rm tr}$, we can
rewrite $\zeta(s)$ as an integral,
\begin{equation}
\zeta(s) = \frac{1}{2\pi i}\int_C \frac{dz}{z^s}
\frac{d}{dz}{\rm tr}\ln u(\tau_+|m^2-z),
\label{contour}
\end{equation}
over the contour $C$ in the complex plane of $z$, which encircles
all roots of (\ref{det}).

It is necessary to note that positivity and real-valuedness of
roots of (\ref{det}) is guaranteed by self-adjointness
and positive definiteness of  $F(d/d\tau) +m^2a$, which is assumed here.

It should be stressed also that each basis function may be taken
with an overall normalization factor depending
on $(m^2 - z)$. This can lead to the additional roots of Eq.
(\ref{det}), which are irrelevant to the eigenvalues of
the elliptic operator under consideration. 
To avoid such an effect, consider the singular point of
the radial equation at $\tau = 0$. The
asymptotic behaviour of $u_A^i(\tau|m^2-z)$ for $\tau \rightarrow 0$ has,
according to the asymptotic expansion theory, a power-law form
\cite{Olver}
\begin{equation}
u(\tau|m^2-z) \sim u_0\tau^k + O(\tau^{k+1}), \ \ \tau \rightarrow 0,
\label{asymp}
\end{equation}
where $k$ is a positive integer number. Thus, to avoid the additional
roots of Eq. (\ref{det}), it is sufficient to require that
$u_{0}$ should be independent of the mass.

If we now assume that the basis functions are analytic in the complex
plane of the mass variable $m^2$, then we can continuously
deform the original contour of integration $C$
to the new contour $\tilde{C}$, which encircles the cut in the complex
plane of the functions $z^{-s}$, coinciding with
the negative real axis. Thus, the general expression for $\zeta(s)$,
to be analytically continued to $s=0$, looks like
\begin{equation}
\zeta(s)=\frac{1}{2\pi i}\int_{\tilde{C}}
\frac{dz}{z^s}\frac{d}{dz}{\rm tr}\ln u(\tau_+|m^2-z).
\label{contour1}
\end{equation}
For the analytic continuation of (\ref{contour1}) from the convergence
region domain to $s = 0$, take into account that the contour
$\tilde{C}$ includes the two boundaries of the negative real axis and a
circle around the point $z = 0$ of some small radius
$\varepsilon$. Therefore
\begin{eqnarray}
&&\zeta(s) = \frac{\sin (\pi s)}{\pi}\int_0^{\infty}
\frac{dM^2}{M^{2s}}\frac{d}{dM^2}{\rm tr}\ln u(\tau_+|m^2+M^2)\nonumber \\
&&+\frac{1}{2\pi i}\int_{C_{\varepsilon}} \frac{dz}{z^s}
\frac{d}{dz}{\rm tr}\ln u(\tau_+|m^2-z),
\label{contour2}
\end{eqnarray}
where the first term is a jump of the integrand in (\ref{contour1})
on the cut of the function $z^{-s}$, integrated along this cut $z = -M^2$.

Let us transform Eq. (\ref{contour2}) by the following sequence of
operations: first analytically continue both terms into the neighborhood
of $s = 0$ and then go to the limit $\varepsilon = 0$. The integral
along $C_{\varepsilon}$ will vanish because of the regularity of
$u(\tau_+|m^2-z)$ at $z = 0$.

It is not so difficult to show that for a quantum-mechanical system with
a finite number of degrees of freedom, as $s\to 0$ we have
\begin{equation}
\zeta(s) = I_{\rm log} + s[I]^{\infty}_0 + O(s^2),
\label{zeta100}
\end{equation}
where
\begin{equation}
I(M^2) \equiv {\rm tr} \ln u(\tau_+|m^2+M^2),
\label{IM}
\end{equation}
$I_{\rm log}$ is the coefficient of $\ln M^2$ in the expansion of
$I$ as $M^2 \rightarrow \infty$, $[I]^{\infty}$ is the
regular part of this expansion as $M^2 \rightarrow \infty$
and $[I]_0 = I(0), [I]^{\infty}_0=[I]^{\infty} - [I]_0$.
It is obvious that in this case
\begin{equation}
\zeta(0) = I_{\rm log},
\label{Ilog}
\end{equation}
 \begin{equation}
 \zeta'(0) = [I]^{\infty}_0.
\label{zetaprime}
 \end{equation}
This result shows that $\zeta(0)$ and $\zeta'(0)$ get a contribution
from the asymptotic value of the basis fucntion
$u_A^i(\tau|M^2)$ for $M^2 \rightarrow \infty$. But their asymptotic
behaviour can be obtained from the JWKB approximation
for the corresponding equation \cite{Olver}.

The problem becomes much more complicated when we study field theories,
which have an infinite number of modes, because the trace
in (\ref{contour1}) becomes the divergent series
\begin{equation}
\zeta(s)=\frac{1}{2\pi i}\int_{\tilde{C}} \frac{dz}{z^s}
\frac{d}{dz}\sum_A[\ln u(\tau_+|m^2-z)]_A^A.
\label{contour3}
\end{equation}
The question arises of how the parameter $s$ can regularize this
divergent series. Let us interchange the summation and integration
operations in (\ref{contour3}),
\begin{equation}
\zeta(s)=\frac{1}{2\pi i}\sum_A\int_{\tilde{C}}
\frac{dz}{z^s}\frac{d}{dz}[\ln u(\tau_+|m^2-z)]_A^A,
\label{contour4}
\end{equation}
and consider the asymptotic behaviour of the integral
\begin{equation}
\int_{\tilde{C}} \frac{dz}{z^s}\frac{d}{dz}[\ln u(\tau_+|m^2-z)]_A^A,
\label{contour5}
\end{equation}
for the collective index $A$ growing to infinity. The numerical
parameter tending to infinity with the growth of $A$
is the parameter $n$ enumerating the harmonics of the radial equation.
Thus, the question of convergence for the sum (\ref{contour4})
reduces to the analysis of the asymptotic behaviour of
(\ref{contour5}) for $n \rightarrow \infty$. Fortunately, the so-called
uniform JWKB expansion for the basis functions has an important
property \cite{Olver,Thorne,BKK1992}: when it is considered as a function
of the two arguments $n \rightarrow \infty$ and the ratio $z/n^2$,
\begin{equation}
\ln u(\tau|m^2-z) = \varphi_{\rm JWKB}\left(n^2,\frac{z}{n^2}\right),
\label{unif}
\end{equation}
then it is uniform in the second argument, $0 \leq |z|/n^2 < \infty$, and
at most has a power-law growth of finite order $k$ in the first argument,
$n^2 \rightarrow \infty$. Therefore, substituting (\ref{unif})
into (\ref{contour5}) and making the change of integration variable
$z \rightarrow n^2z$ one finds that this integral has an asymptotic behaviour,
\begin{equation}
\frac{1}{n^{2s}}\int_{\tilde{C}}\frac{dz}{z^s}\frac{d}{dz}
\varphi_{\rm JWKB}(n^2,z),
\label{WKB}
\end{equation}
which converges for some $ s > 0$ and, due to uniformity of
$\varphi_{\rm JWKB}(n^2,z)$
in $z$, has a bound const $\times (n^2)^{k-s}$ providing the
convergence of the infinite
series (\ref{contour4}) for some large positive $s$. Thus, large
values of $s > 0$ regularize the divergent sum in (\ref{contour3}).

Making the change of integration variable $z \rightarrow n^2z$ in
Eq. (\ref{contour3}) and
interchanging back the order of integration and summation one can
represent the $\zeta$ function in the form
\begin{equation}
\zeta(s) = \frac{1}{2\pi i}\int_{\tilde{C}}\frac{dz}{z^s}
\frac{d}{dz}I(-z,s),
\label{contour6}
\end{equation}
where $I(-z,s)$ is the manisfestly regularized infinite sum
\begin{equation}
 I(-z,s) = \sum_A\frac{1}{n^{2s}}[\ln u(\tau_+|m^2-z)]_A^A.
 \label{inf-sum}
 \end{equation}

Similarly to (\ref{contour2}), we can split the integral (\ref{contour6})
over $\tilde{C}$ into a sum of two terms and show that the
integral around the circle $C_{\varepsilon}$ tends to zero.
However, unlike models with a finite number of physical
variables, the series (\ref{inf-sum}) analytically continued from its
convergence domain generally has a pole at $s = 0$,
 \begin{equation}
 I(M^2,s) = \frac{I^{\rm pole}(M^2)}{s} + I^R(M^2)  +  O(s).
 \label{infinitedeg}
 \end{equation}
Therefore, instead of the formula (\ref{zeta100}) we obtain the
following result for the case of field theory:
\begin{eqnarray}
&&\zeta(s) = (I^R)_{\rm log} + [I^{\rm pole}]_0^{\infty}\nonumber \\
&&+\left\{[I^R]_0^{\infty}-\int_0^{\infty} dM^2 \ln M^2
\frac{dI^{\rm pole}(M^2)}{dM^2}
\right\} + O(s^2),
\label{zeta101}
\end{eqnarray}
where $()_{\rm log}$ and $[]_0^{\infty}$ have the same sense as in the
quantum-mechanical case. Thus we have
\begin{equation}
\zeta(0) =  (I^R)_{\rm log} + [I^{\rm pole}]_0^{\infty},
\label{zeta102}
\end{equation}
\begin{equation}
\zeta'(0) = [I^R]_0^{\infty}-\int_0^{\infty} dM^2 \ln M^2
\frac{dI^{\rm pole}(M^2)}{dM^2}.
\label{zeta103}
\end{equation}

These equations generalize the algorithms (\ref{Ilog}) and
(\ref{zetaprime}) to field theories with an infinite number of physical
modes. But these generalizations are non-trivial: only the terms
$(I^R)_{\rm log}$ in (\ref{zeta102}) and
$[I^R]_0^{\infty}$ in (\ref{zeta103}) are similar to the
expressions (\ref{Ilog}) and (\ref{zetaprime}). The terms including
$I^{\rm pole}$ do not have analogs in a theory with a finite
number of modes. These terms are responsible for the
non-trivial renormalization of the ultraviolet divergences
performed by the $\zeta$-function regularization.

The formalism just described is a fine tuned scheme for the computation of
$\zeta (0)$ and $\zeta ' (0)$. It does, however, not allow to extract 
other properties of $\zeta (s)$. But
in order to determine heat kernel coefficients or the Casimir energy 
associated with some
quantum field theory models, other particular properties are 
needed \cite{K2001,eliz95b,bord09b}. These can be found by
analytically continuing the zeta function $\zeta (s)$ as given in 
Eq. (\ref{contour4}) to a meromorphic function in
the complex plane. The details of this procedure depend very much on the 
explicit form of $u (\tau _+| m^2 -z)$ in Eq. (\ref{contour4}).
In general one can only say that adding and subtracting the asymptotic 
expansion briefly outlined in Eq. (\ref{unif}) is crucial to the method, 
but the precise nature
of integrals and series to be done to obtain the analytical continuation 
depend on exactly what $\varphi_{JWKB} (n^2, z/n^2)$ actually is. 
For the example of the scalar
Laplacian on a four dimensional ball with various boundary conditions 
details will be provided in the next section.

\section{Dirichlet, Neumann and Robin Boundary Conditions}

One becomes familiar with Dirichlet boundary conditions as
soon as one studies potential theory. The first boundary-value
problem of potential theory is the existence of a function,
harmonic in a closed region, and taking on preassigned
continuous boundary values. This is known as the Dirichlet
problem, and is the oldest existence theorem in potential
theory. Usually, one first tries to express a harmonic function
in terms of its boundary values. One then sees if the expression
found continues to represent a harmonic function when the
boundary values are any given continuous function.

The problem of finding a function, harmonic in a region, and
having normal derivatives equal to the function given on the
boundary is instead the Neumann problem, or the second
boundary-value problem of potential theory. The theorem asserting
the existence of a solution of this problem is known as the
second fundamental existence theorem of potential theory
\cite{K1954}.

In the semiclassical approximation of the quantum theory of
a real scalar field in a real Riemannian background with
boundary, the guiding principle for the
choice of boundary conditions is that the boundary data should
reflect those particular conditions which lead to a well-posed
classical boundary-value problem. Thus, on using the
background-field method, the scalar-field perturbations
$\varphi$ are required to obey one of the following three
boundary conditions on the bounding surfaces \cite{K1978}:
\vskip 0.3cm
\noindent
(i) Dirichlet problem:
\begin{equation}
\varphi=0 \; \; \; \; {\rm at} \; \; \; \;
{\partial M},
\label{(6.1.1)}
\end{equation}
\vskip 0.3cm
\noindent
(ii) Neumann problem:
\begin{equation}
{\partial \varphi \over \partial \tau}=0
\; \; \; \; {\rm at} \; \; \; \;
{\partial M},
\label{(6.1.2)}
\end{equation}
\vskip 0.3cm
\noindent
(iii) Robin problem:
\begin{equation}
{\partial \varphi \over \partial \tau}
+{u\over \tau} \varphi =0
\; \; \; \; {\rm at} \; \; \; \;
{\partial M}.
\label{(6.1.3)}
\end{equation}
For example, in the case of a massless scalar field at
1 loop about flat 4-dimensional Euclidean space
bounded by a 3-sphere, the technique of section 2
may be used to find the following values for
the resulting anomalous scaling factors:
\begin{equation}
\zeta_{D}(0)=-{1\over 180}
\; \; \; \; ({\rm Dirichlet}),
\label{(6.1.4)}
\end{equation}
\begin{equation}
\zeta_{N}(0)={29\over 180}
\; \; \; \; ({\rm Neumann}),
\label{(6.1.5)}
\end{equation}
\begin{equation}
\zeta_{R}(0)= -{1\over 180}-{1\over 6}(u-1)^{3}
\; \; \; \; ({\rm Robin}).
\label{(6.1.6)}
\end{equation}
In this particular case, the $\zeta(0)$ values coincide with
the conformal anomaly, since massless scalar field theories
are conformally invariant in flat space-time.
It was not until in \cite{M1989} that a powerful {\it analytic}
algorithm was developed for the analysis of the Robin
case, and the first correct {\it geometric} results for
$\zeta(0)$ were only published in \cite{MD1989}
and \cite{V1995,MP1994}. More recent
work on real scalar fields on the Euclidean ball in various
dimensions can be found in \cite{bord96-37-895,BGKE1996,dowk96-13-585,
dowk96-366-89,bord96-182-371}.

In order to outline the contour integration method for the analysis of 
spectral zeta functions as a function of $s$
let us exploit this opportunity and rederive Eq. (\ref{(6.1.4)}) with an 
indication on how to obtain Eqs. (\ref{(6.1.5)})-(\ref{(6.1.6)}). 
Given the treatment of arbitrary dimension
does not cause any additional complications, we will consider the 
$D=d+1$ dimensional ball \cite{K2001}.

A massless scalar field leads to the eigenvalue problem for a Laplacian,  
and for the spherically symmetric problem at hand the use
of polar coordinates seems appropriate. In these coordinates the 
Laplacian reads
\begin{eqnarray}
\Delta = \frac{\partial^2}{\partial r^2} + \frac d r \frac 
\partial {\partial r} + \frac 1 {r^2} \Delta _{{\cal N}}, \label{kk1}
\end{eqnarray}
with $\Delta_{{\cal N}}$ the Laplacian on the $d$-dimensional sphere, 
${\cal N} = S^d$.

By imposing Dirichlet boundary conditions on the sphere, the boundary of 
the ball, eigenvalues are determined by the transcendental equation
\begin{eqnarray}
J_\nu (\lambda ) =0, \label{kk2}
\end{eqnarray}
with $\nu = \ell + (d-1)/2$, $\ell =0,1,2,...$, and with the radius $R$ 
of the ball being chosen as $R=1$. The degeneracy $d_\nu$ of each eigenvalue
equals the degeneracy of the eigenvalues of the Laplacian on the 
$d$-sphere and for $d \geq 2$, the case we will concentrate on in the 
following, it equals
\begin{eqnarray}
d_\nu = (2\ell +d-1) \frac{(\ell + d-2)!}{\ell ! (d-1)!} .
\end{eqnarray}
The information provided suffices to give an explicit representation of 
the associated zeta function as given in Eq. (\ref{contour4}), i.e.
\begin{eqnarray}
\zeta (s) &=& \frac 1 {2\pi i} \sum_\nu d_\nu \int
\limits_{\tilde C} \frac {dz} {z^{2s}} \frac d {dz} 
\ln \left( z^{-\nu} J_\nu (z) \right) \nonumber\\
&=& \frac {\sin \pi s} \pi \sum_\nu d_\nu \int
\limits_0^\infty \frac{dk} {k^{2s}} \frac d {dk} 
\ln \left( k^{-\nu} I_\nu (k) \right) , \label{kk3}
\end{eqnarray}
where this equality is obtained by deforming the contour $\tilde C$ 
to the imaginary axis. The relevant uniform asymptotic behaviour outlined 
in Eq. (\ref{unif}) for the given example follows from
\begin{eqnarray} 
I_\nu (\nu z) & \sim & \frac 1 {\sqrt {2\pi \nu}} 
\frac{{\rm e}^{\nu \eta}}{(1+z^2)^{1/4}} \left[ 1 
+ \sum_{k=1}^\infty \frac{u_k (t)} {\nu^k}\right], \label{kk4}
\end{eqnarray}
valid for $\nu\to \infty$ as $z=k/\nu$ is fixed \cite{Olver,abra70b}. 
Here $t=1/\sqrt{1+z^2}$ and $\eta = \sqrt{1+z^2} 
+ \ln [z/(1+\sqrt{1+z^2})]$. Higher powers in $t$ follow
from the recursion \cite{abra70b}
\begin{eqnarray}
u_{k+1} (t) = \frac 1 2 t^2 (1-t^2) u_k ' (t) + \frac 1 8 
\int\limits_0^t d\tau (1-5\tau^2) u_k (\tau ) ,\nonumber
\end{eqnarray}
starting with $u_0 (t) =1$. 
On defining polynomials $D_n (t)$ from the expansion
\begin{eqnarray}
\ln \left[ 1+\sum_{k=1}^\infty \frac{u_k (t)} {\nu^k} \right] 
\sim \sum_{n=1}^\infty \frac{D_n (t)} {\nu^n} , \label{kkd}
\end{eqnarray}
the leading few polynomials are
\begin{eqnarray}
D_1 (t) &=& \frac 1 8 t - \frac 5 {24} t^3, \nonumber\\
D_2 (t) &=& \frac 1 {16} t^2 - \frac 3 8 t^4 
+ \frac 5 {16} t^6, \label{kkd1}\\
D_3 (t) &=& \frac {25} {384} t^3 - \frac{531}{640} t^5 
+ \frac{ 221} {128} t^7 - \frac{1105}{1152} t^9,\nonumber
\end{eqnarray}
with many more polynomials easily found 
by using an algebraic computer program.

By adding and subtracting $N$ leading terms, for $\zeta (0)$ in $D=4$ 
we will ultimately choose $N=3$, the zeta function (\ref{kk3}) splits 
into the pieces (after substituting $k=z\nu$)
\begin{eqnarray}
\zeta (s) = Z(s) + \sum_{i=-1}^N A_i (s) , \nonumber
\end{eqnarray}
where
\begin{eqnarray}
Z (s) &=& \frac{ \sin (\pi s)} \pi \sum_\nu \,\, d_\nu \,\, \int
\limits_0^\infty dz (z\nu)^{-2s} \frac \partial {\partial z} 
\left\{ \ln \left[ z^{-\nu} I_\nu (z\nu) \right] \right.\nonumber\\
& & \left.- \ln \left[ \frac{z^{-\nu}} {\sqrt{2\pi \nu}} 
\frac{{\rm e}^{\nu \eta}}{(1+z^2)^{1/4}} \right] - \sum_{n=1}^N 
\frac{D_n (t)} {\nu^n}\right\} ,\label{kkz}
\end{eqnarray}
and the $A_i (s)$ result from the different orders in the asymptotic 
expansion, explicitly
\begin{eqnarray}
A_{-1} (s) &=& \frac{\sin (\pi s)} \pi \sum_\nu \,\, d_\nu \,\,
\int\limits_0^\infty dz (z\nu)^{-2s} \frac \partial {\partial z} 
\ln \left( z^{-\nu } {\rm e}^{\nu \eta} \right), \nonumber\\
A_0 (s) &=& \frac{\sin (\pi s)} \pi \sum_\nu \,\, d_\nu \,\,
\int\limits_0^\infty dz (z\nu)^{-2s} \frac \partial {\partial z} 
\ln (1+z^2)^{-1/4} , \nonumber\\
A_i (s) &=& \frac{\sin (\pi s)} \pi \sum_\nu \,\, d_\nu \,\,
\int\limits_0^\infty dz (z\nu)^{-2s} \frac \partial {\partial z} 
\left( \frac{ D_i (t)} {\nu^i} \right) .\nonumber
\end{eqnarray}
It can be shown that $Z(s)$ is analytic in the half-plane 
$(d-1-N)/2<\Re s$. These formulas therefore make it possible 
to find a representation 
of $\zeta (s)$ valid for any value of $s$.

Choosing $N$ suitably large, given the factor $\sin (\pi s)$ in 
(\ref{kkz}), $Z(s)$ will therefore not contribute to $\zeta (0)$.
As far as $\zeta (0)$ is concerned, it therefore suffices to only consider 
the $A_i (s)$ further. By introducing the so-called base zeta function,
\begin{eqnarray}
\zeta _{{\cal N}} (s) = \sum_\nu d_\nu \,\, \nu^{-2s}, \label{kk5}
\end{eqnarray}
$A_{-1}(s)$ and $A_0 (s)$ are readily evaluated as
\begin{eqnarray}
A_{-1} (s) &=& \frac 1 {4\sqrt \pi} \frac{\Gamma \left( s- \frac 1 2 
\right)} {\Gamma (s+1)} \zeta_{{\cal N}} \left( s-\frac 1 2 \right) , 
\nonumber\\
A_0 (s) &=& - \frac 1 4 \zeta _{{\cal N}} (s) .\nonumber
\end{eqnarray}
In order to compute the higher $A_i (s)$ note that the polynomials 
$D_i (t)$ can be written as
\begin{eqnarray}
D_i (t) = \sum_{b=0}^i x_{i,b}\,\, t^{i+2b} , \nonumber
\end{eqnarray}
with the coefficients $x_{i,b}$ easily determined from the definition 
(\ref{kkd}) of $D_i (t)$; to read off the numbers $x_{i,b}$ for $i=1,2,3$, 
see also (\ref{kkd1}).
The $z$-integrals are then easily done and one finds
\begin{eqnarray}
A_i (s) &=& - \frac 1 {\Gamma (s)} \zeta_{{\cal N}} 
\left( s+\frac i 2 \right) \sum_{b=0}^i x_{i,b} 
\frac{\Gamma \left( s+b+\frac i 2 \right)} {\Gamma \left( 
b+\frac i 2 \right)} .\nonumber
\end{eqnarray}
Concentrating on $\zeta (0)$ in four dimensions we note that
\begin{eqnarray}
\zeta_{{\cal N}} (s) = \sum_{\ell =0}^\infty (\ell +1)^2 
\,\, (\ell +1 )^{-2s} = \zeta_R (2 s-2),\nonumber
\end{eqnarray}
which allows us to write
\begin{eqnarray}
A_{-1} (s) &=& \frac 1 {4 \sqrt \pi} \frac{\Gamma \left( 
s-\frac 1 2 \right)}{\Gamma (s+1)} \zeta_R(2s-3) , \nonumber\\
A_0 (s) &=& - \frac 1 4 \zeta_R (2s-2) , \nonumber\\
A_i (s) &=&- \frac 1 {\Gamma (s)} \zeta_R (2s+i-2) 
\sum_{b=0}^i x_{i,b} \frac{\Gamma \left( s+b+\frac i 2\right)}
{\Gamma \left( b+\frac i 2 \right)}.\nonumber
\end{eqnarray}
At $s=0$ we compute
\begin{eqnarray}
A_{-1} (0) &=& \frac 1 {4\sqrt \pi} \Gamma \left( - 
\frac 1 2 \right) \zeta _R (-3) = - \frac 1 {240} , \nonumber\\
A_0 (0) &=& A_1 (0) = A_2 (0) =0, \nonumber\\
A_3 (0) &=& - \frac 1 2 \sum_{b=0}^3 x_{3,b} = - \frac 1 {720}, \nonumber
\end{eqnarray}
and thus as stated
\begin{eqnarray}
\zeta (0) = - \frac 1 {240} - \frac 1 {720} = - \frac 1 {180}.\nonumber
\end{eqnarray}
In the same manner Neumann and Robin boundary conditions can be 
treated by starting with the implicit eigenvalue equation
\begin{eqnarray}
\left( 1 - \frac D 2 +u\right) J_\nu (\lambda ) + \lambda 
J_\nu ' (\lambda ) =0,\nonumber
\end{eqnarray}
where $u=0$ corresponds to Neumann boundary conditions; for details about 
the very similar calculations we refer to \cite{K2001}.

For complex scalar fields, the most general case corresponds
to mixed boundary conditions, i.e. when the real part obeys
Dirichlet conditions and the imaginary part obeys Neumann
conditions, or the other way around.
In the light of (\ref{(6.1.4)})--(\ref{(6.1.5)}), the resulting
conformal anomaly for a complex massless field on the Euclidean
ball is found to be
\begin{equation}
\zeta(0)={7\over 45} .
\label{(6.1.7)}
\end{equation}

\section{Mixed boundary conditions for gauge fields}

We are interested in the 1-loop amplitudes of vacuum
Maxwell theory in the presence of boundaries. Since in the
classical theory the potential $A_{\mu}$ is subject to the
gauge transformations
\begin{equation}
{\widehat A}_{\mu} \equiv A_{\mu}+\partial_{\mu}\varphi,
\label{(6.4.1)}
\end{equation}
this gauge freedom is reflected in the quantum theory by a
ghost 0-form, i.e. an anticommuting, complex
scalar field, hereafter denoted again by $\varphi$.
The two sets of mixed boundary conditions
consistent with gauge invariance and Becchi--Rouet--Stora--Tyutin
(hereafter BRST) symmetry are magnetic, i.e.
\begin{equation}
\Bigr[A_{k}\Bigr]_{\partial M}=0,
\label{(6.4.2a)}
\end{equation}
\begin{equation}
\Bigr[\Phi(A)\Bigr]_{\partial M}=0,
\label{(6.4.2b)}
\end{equation}
\begin{equation}
\Bigr[\varphi\Bigr]_{\partial M}=0,
\label{(6.4.2c)}
\end{equation}
or electric, i.e.
\begin{equation}
\Bigr[A_{0}\Bigr]_{\partial M}=0,
\label{(6.4.3a)}
\end{equation}
\begin{equation}
\left[{\partial A_{k}\over \partial \tau}\right]_{\partial M}
=0,
\label{(6.4.3b)}
\end{equation}
\begin{equation}
\left[{\partial \varphi \over \partial \tau}\right]_{\partial M}
=0,
\label{(6.4.3c)}
\end{equation}
where $\Phi$ is an arbitrary gauge-averaging functional defined
on the space of connection 1-forms $A_{\mu}dx^{\mu}$. Note
that the boundary condition (\ref{(6.4.2c)})
ensures the gauge invariance of the boundary conditions
(\ref{(6.4.2a)})-(\ref{(6.4.2b)}) on making the gauge
transformation (\ref{(6.4.1)}).
Similarly, the boundary condition (\ref{(6.4.3c)})
ensures the gauge invariance of
(\ref{(6.4.3a)})-(\ref{(6.4.3b)}) on transforming
the potential as in (\ref{(6.4.1)}).
For example, when the Lorenz
gauge-averaging functional is chosen,
$$
\Phi_{L}(A) \equiv \nabla^{\mu}A_{\mu},
$$
the boundary condition (\ref{(6.4.2b)}) reduces to ($K$ being the
extrinsic-curvature tensor of the boundary)
$$
\biggr[{\partial A_{0}\over \partial \tau}
+A_{0}{\rm Tr} \; K \biggr]_{\partial M}=0,
$$
by virtue of Eq. (\ref{(6.4.2a)}) that sets to zero at the boundary all
longitudinal and transverse modes.

It is also instructive to prove the BRST invariance of our
boundary conditions. For this purpose, take e.g. the electric
boundary conditions (\ref{(6.4.3a)})--(\ref{(6.4.3c)}),
jointly with the BRST transformations (the ghost 0-form
corresponding to independent and real-valued ghost fields
denoted by $\omega$ and $\psi$, while $\delta \lambda$ is
an anti-commuting gauge parameter)
\begin{equation}
\delta_{\rm BRST} A_{\mu}=(\nabla_{\mu}\psi)\delta \lambda,
\label{(x1)}
\end{equation}
\begin{equation}
\delta_{\rm BRST} \omega=(\nabla^{\mu}A_{\mu})\delta \lambda,
\label{(x2)}
\end{equation}
\begin{equation}
\delta_{\rm BRST} \psi=0.
\label{(x3)}
\end{equation}
Now if $\psi$ obeys Neumann boundary conditions,
\begin{equation}
\Bigr[n_{\mu}\nabla^{\mu}\psi \Bigr]_{\partial M}=0,
\label{(6.4.4)}
\end{equation}
then by virtue of (\ref{(x1)}) one finds (hereafter
${\widehat \delta} \equiv \delta_{\rm BRST}$)
\begin{equation}
{\widehat \delta}\Bigr(n_{\mu}A^{\mu}\Bigr)
=n_{\mu}\Bigr({\widehat \delta}A^{\mu}\Bigr)
=(\delta \lambda)n_{\mu}\Bigr(\nabla^{\mu} \psi \Bigr),
\label{(6.4.5)}
\end{equation}
and this variation vanishes at the boundary by virtue of
(\ref{(6.4.4)}). Thus, the boundary condition
\begin{equation}
\Bigr[n_{\mu}A^{\mu}\Bigr]_{\partial M}=0,
\label{(6.4.6)}
\end{equation}
which is the covariant form of (\ref{(6.4.3a)}), is preserved under
the action of BRST transformations. Further details can be
found in \cite{MP1990}.

For a given choice of one of these two sets of mixed boundary
conditions, different choices of background 4-geometry,
boundary 3-geometry and gauge-averaging functional lead
to a number of interesting results.
We here summarize them in the
case of a background given by flat Euclidean
4-space bounded by one 3-sphere (i.e. the disk) or by
two concentric 3-spheres (i.e. the ring).
\vskip 0.3cm
\noindent
(i) The operator matrix acting on normal and longitudinal
modes of the potential can be put in diagonal form for all relativistic
gauge conditions which can be expressed as
\begin{equation}
\Phi_{b}(A) \equiv \nabla^{\mu}A_{\mu}-b \; A_{0} \;
{\rm Tr} \; K,
\label{(6.4.7)}
\end{equation}
where $\nabla^{\mu}$ denotes covariant differentiation
with respect to the Levi-Civita connection of
the background, and $b$ is a dimensionless parameter.
\vskip 0.3cm
\noindent
(ii) In the case of the disk, the Lorenz gauge (set $b=0$
in (\ref{(6.4.7)})) leads to a $\zeta(0)$ value
\begin{equation}
\zeta_{L}(0)=-{31\over 90},
\label{(6.4.8)}
\end{equation}
for both magnetic and electric boundary conditions, which agrees
\cite{MP1994} with the geometric theory of the asymptotic heat
kernel. However, the $\zeta(0)$ value depends on the gauge
condition, and unless
$b$ vanishes it also depends on the boundary conditions.
\vskip 0.3cm
\noindent
(iii) In the case of the ring, one finds
\begin{equation}
\zeta(0)=0,
\label{(6.4.9)}
\end{equation}
for all gauge conditions, independently of boundary conditions.
This result agrees with the geometric formulae for the
heat kernel, since volume (i.e. interior)
contributions to $\zeta(0)$ vanish in a flat
background, while surface (i.e. boundary) contributions cancel
each other.
\vskip 0.3cm
\noindent
(iv) In the case of boundary 3-geometries given by one
or two 3-spheres, the most general gauge-averaging
functional takes the form \cite{EKMP95a,EKP1997}
\begin{equation}
\Phi(A)=\gamma_{1} \; { }^{(4)}\nabla^{0}A_{0}
+{\gamma_{2}\over 3}A_{0} \; {\rm Tr} \; K
-\gamma_{3} \; { }^{(3)}\nabla^{i}A_{i},
\label{(6.4.10)}
\end{equation}
where $\gamma_{1},\gamma_{2}$ and $\gamma_{3}$ are arbitrary
dimensionless parameters, which give different ``weight'' 
to the various terms in the 3+1 decomposition of 
$\nabla^{\mu}A_{\mu}$. Thus, unless $\gamma_{1},\gamma_{2}$
and $\gamma_{3}$ take some special values
(cf. (\ref{(6.4.7)})), it is
not possible to diagonalize the operator matrix acting on
normal and longitudinal modes of the potential.
\vskip 0.3cm
\noindent
(v) The contributions to $\zeta(0)$ resulting from normal
and longitudinal modes {\it do not} cancel in general the
contribution of ghost modes. Thus,
{\it transverse modes do not provide the only surviving contribution
to 1-loop amplitudes}. In other words, {\it all} perturbative
modes are necessary to recover the correct form of 1-loop
semiclassical amplitudes.

\section{Boundary conditions for the gravitational field}

For gauge fields and gravitation, the boundary conditions
are mixed in that some components of the field (more
precisely, a 1-form or a symmetric tensor of type
$(0,2)$) obey a set of boundary
conditions, and the remaining part of the field obeys another
set of boundary conditions. Moreover, the boundary conditions
are invariant under local gauge transformations provided that
suitable boundary conditions are imposed on the corresponding
ghost 0-form or 1-form.

We are here interested in the derivation of mixed boundary
conditions for Euclidean quantum gravity. The knowledge of
the classical variational problem, and the principle of
gauge invariance, are enough to lead to a highly non-trivial
quantum boundary-value problem. Indeed, it is by now well-known
that, if one fixes the 3-metric at the boundary in
general relativity, the corresponding variational problem
is well-posed and leads to the Einstein equations, providing
the Einstein-Hilbert action is supplemented by a boundary
term whose integrand is proportional to the trace of the
second fundamental form. In the corresponding
quantum boundary-value problem, which is relevant for the
1-loop approximation in quantum gravity, the perturbations
$h_{ij}$ of the induced 3-metric are set to zero at the
boundary. Moreover, the whole set of metric perturbations
$h_{\mu \nu}$ are subject to the so-called infinitesimal
{\it gauge} transformations
\begin{equation}
{\widehat h}_{\mu \nu} \equiv h_{\mu \nu}
+\nabla_{(\mu} \; \varphi_{\nu)},
\label{(6.5.1)}
\end{equation}
where $\nabla$ is the Levi-Civita connection of the background
4-geometry with metric $g$, and $\varphi_{\nu}dx^{\nu}$
is the ghost 1-form. In geometric language, the
difference between ${\widehat h}_{\mu \nu}$ and
$h_{\mu \nu}$ is given by the Lie derivative along $\varphi$
of the 4-metric $g$.

For problems with boundary, Eq. (\ref{(6.5.1)}) implies that
\begin{equation}
{\widehat h}_{ij}=h_{ij}+\varphi_{(i \mid j)}
+K_{ij} \varphi_{0},
\label{(6.5.2)}
\end{equation}
where the stroke denotes, as usual, 3-dimensional
covariant differentiation tangentially with respect to the
intrinsic Levi-Civita connection of the boundary, while
$K_{ij}$ is the extrinsic-curvature tensor of the boundary.
Of course, $\varphi_{0}$ and $\varphi_{i}$ are the normal
and tangential components of the ghost 1-form, respectively.
Note that boundaries make it necessary to perform a 3+1
split of space-time geometry and physical fields.
As such, they introduce non-covariant elements in the analysis
of problems relevant for quantum gravity. This seems to be
an unavoidable feature, although the boundary conditions may
be written in a covariant way.

In light of (\ref{(6.5.2)}), the boundary conditions
\begin{equation}
\Bigr[h_{ij}\Bigr]_{\partial M}=0
\label{(6.5.3a)}
\end{equation}
are gauge invariant, i.e.
\begin{equation}
\Bigr[{\widehat h}_{ij}\Bigr]_{\partial M}=0,
\label{(6.5.3b)}
\end{equation}
if and only if the whole ghost 1-form obeys homogeneous
Dirichlet conditions, so that
\begin{equation}
\Bigr[\varphi_{0}\Bigr]_{\partial M}=0,
\label{(6.5.4)}
\end{equation}
\begin{equation}
\Bigr[\varphi_{i}\Bigr]_{\partial M}=0.
\label{(6.5.5)}
\end{equation}
The conditions (\ref{(6.5.4)}) and (\ref{(6.5.5)})
are necessary and sufficient
since $\varphi_{0}$ and $\varphi_{i}$ are independent, and
3-dimensional covariant differentiation commutes with the
operation of restriction at the boundary. Indeed, we are
assuming that the boundary is smooth and not totally geodesic,
i.e. $K_{ij} \not = 0$. However, {\it at those points of} $\partial M$
{\it where the extrinsic-curvature tensor vanishes},
{\it the condition} (\ref{(6.5.4)}) {\it is no longer necessary}
\cite{EKP1997}.

The problem now arises to impose boundary conditions on the
remaining set of metric perturbations. The key point is to
make sure that the invariance of such boundary conditions under
the infinitesimal transformations (\ref{(6.5.1)}) is again
guaranteed by (\ref{(6.5.4)})-(\ref{(6.5.5)}),
since otherwise one would obtain incompatible sets of
boundary conditions on the ghost 1-form. Indeed, on using
the Faddeev--Popov formalism for the amplitudes of quantum
gravity, it is necessary to use a gauge-averaging term in
the Euclidean action, of the form
\begin{equation}
I_{\rm g.a.} \equiv {1\over 32 \pi G \alpha}
\int_{M}\Phi^{\mu}g_{\mu \nu}\Phi^{\nu}
\sqrt{{\rm det} \; g} \; d^{4}x,
\label{(6.5.6)}
\end{equation}
where $\Phi^{\mu}$ is any relativistic gauge-averaging
functional which leads to self-adjoint elliptic operators
on metric and ghost perturbations. One then finds that
(here $\Phi_{\mu} \equiv g_{\mu \nu}\Phi^{\nu}$)
\begin{equation}
\delta \Phi_{\mu}(h) \equiv
\Phi_{\mu}(h)-\Phi_{\mu}(\widehat h)
={\cal F}_{\mu}^{\; \; \nu} \; \varphi_{\nu},
\label{(6.5.7)}
\end{equation}
where ${\cal F}_{\mu}^{\; \; \nu}$ is an elliptic operator
that acts linearly on the ghost 1-form. Thus, if one
imposes the boundary conditions
\begin{equation}
\Bigr[\Phi_{0}(h)\Bigr]_{\partial M}=0,
\label{(6.5.8a)}
\end{equation}
\begin{equation}
\Bigr[\Phi_{i}(h)\Bigr]_{\partial M}=0,
\label{(6.5.9a)}
\end{equation}
their invariance under (\ref{(6.5.1)})
is guaranteed when (\ref{(6.5.4)}) and (\ref{(6.5.5)})
hold, by virtue of (\ref{(6.5.7)}). Hence one also has
\begin{equation}
\Bigr[\Phi_{0}(\widehat h)\Bigr]_{\partial M}=0,
\label{(6.5.8b)}
\end{equation}
\begin{equation}
\Bigr[\Phi_{i}(\widehat h)\Bigr]_{\partial M}=0.
\label{(6.5.9b)}
\end{equation}

In section 7 we shall study this scheme, first proposed in
\cite{B1987}, when the linear covariant gauge of the de Donder
type is chosen. We will see that this leads to boundary conditions
which involve normal and tangential derivatives of normal
components of metric perturbations, and the resulting 1-loop
divergence will be evaluated.

\section{Equations for basis functions and their solutions
for pure gravity}

For the reasons described in the introduction, we study pure
gravity at 1 loop about a flat Euclidean background with
two concentric 3-sphere boundaries, and eventually let one of
the 3-spheres shrink to a point. Our approach to
quantization follows the Feynman--DeWitt--Faddeev--Popov formalism
\cite{DW2003}. Hence we deal with quantum amplitudes of the form
$$
Z[{\rm boundary \; data}]=\int_{C} \mu_{1}[g] \; \mu_{2}[\varphi]
\; {\rm exp}(-{\widetilde I}_{E}).
$$
With our notation, $C$ is the set of all Riemannian
4-geometries matching the boundary data,
$\mu_{1}$ is a suitable measure on the space
of metrics, $\mu_{2}$ is a suitable measure for ghosts,
$\Phi^{\mu}$ is an arbitrary gauge-averaging functional, and the
total Euclidean action reads (in $c=1$ units)
\begin{eqnarray}
{\widetilde I}_{E}&=&I_{\rm gh}+{1\over 16 \pi G}
\int_{M}{ }^{(4)}R \sqrt{{\rm det} \; g} \; d^{4}x
+{1\over 8 \pi G} \int_{\partial M}{\rm Tr} \; K
\sqrt{{\rm det} \; q} \; d^{3}x \nonumber \\
&+& {1\over 16\pi G} \int_{M}
{1\over 2\alpha} \Phi^{\mu}g_{\mu \nu}\Phi^{\nu}
\sqrt{{\rm det} \; g} \; d^{4}x,
\label{(9.2.1)}
\end{eqnarray}
where ${ }^{(4)}R$ is the trace of the 4-dimensional Ricci tensor.
Of course, $K$ is the extrinsic-curvature tensor of the
boundary, $q$ is the induced 3-metric of $\partial M$,
and $\alpha$ is a positive dimensionless parameter.
The ghost action $I_{\rm gh}$ depends on the specific form of
$\Phi^{\mu}$. Denoting by $h_{\mu \nu}$ the perturbation
about the background 4-metric $g_{\mu \nu}$, one thus finds
field equations of the kind
$$
\cstok{\ }^{\Phi} h_{\mu \nu} = 0 ,
$$
where $\cstok{\ }^{\Phi}$ is the 4-dimensional elliptic operator
corresponding to the form of $\Phi_{\nu} \equiv g_{\nu \mu}\Phi^{\mu}$ 
one is working with. Here we choose the de Donder 
gauge-averaging functional
$$
\Phi_{\nu}^{DD} \equiv
\nabla^{\mu} \left(h_{\mu \nu}-{1\over 2} g_{\mu \nu}
{\hat h} \right) ,
$$
where $\nabla^{\mu}$ is covariant differentiation with respect
to $g_{\mu \nu}$, and ${\hat h} \equiv g^{\mu \nu} \; h_{\mu \nu}$.
The corresponding $\cstok{\ }^{\Phi^{DD}}$ operator is the one obtained
by analytic continuation of the standard D'Alembert operator,
hereafter denoted by $\cstok{\ }$.
The resulting eigenvalue equation is
$$
\cstok{\ } h_{\mu \nu}^{(\lambda)}
+ \lambda h_{\mu \nu}^{(\lambda)} = 0.
$$

Now we can make the 3+1 decomposition of
our background 4-geometry and
expand $h_{00}, h_{0 i}$ and $h_{ij}$ in
hyperspherical harmonics as
\begin{equation}
h_{00}(x,\tau) = \sum_{n=1}^{\infty} a_{n}(\tau) Q^{(n)}(x),
\label{(9.2.2)}
\end{equation}
\begin{equation}
h_{0i}(x,\tau) = \sum_{n=2}^{\infty}
\biggr[b_{n}(\tau)
{Q_{\mid i}^{(n)}(x)\over (n^{2} - 1)} +
c_{n}(\tau) S_{i}^{(n)}(x)\biggr],
\label{(9.2.3)}
\end{equation}
\begin{eqnarray}
h_{ij}(x,\tau)&=& \sum_{n=3}^{\infty}
d_{n}(\tau) \left[{Q_{\mid ij}^{(n)}(x)\over
(n^{2} - 1)} + {c_{ij}\over 3} Q^{(n)}(x)\right]
+\sum_{n=1}^{\infty}{e_{n}(\tau)\over 3} c_{ij} Q^{(n)}(x) \nonumber \\
&+& \sum_{n=3}^{\infty}\biggr[f_{n}(\tau)
\Bigr(S_{i \mid j}^{(n)}(x) + S_{j \mid i}^{(n)}(x)\Bigr)
+k_{n}(\tau) G_{ij}^{(n)}(x)\biggr].
\label{(9.2.4)}
\end{eqnarray}
Here $Q^{(n)}(x), S_{i}^{(n)}(x)$ and $G_{ij}^{(n)}(x)$ are scalar,
transverse vector and transverse-traceless tensor hyperspherical
harmonics, respectively, on a unit 3-sphere with metric $c_{ij}$.

The insertion of the expansions (\ref{(9.2.2)})--(\ref{(9.2.4)})
into Eq.~(\ref{(9.2.1)}) leads to the following
system of equations,
\begin{equation}
{\widehat A}_{n} a_{n}(\tau) + {\widehat B}_{n} b_{n}(\tau)
+ {\widehat C}_{n} e_{n}(\tau) = 0,
\label{(9.2.5)}
\end{equation}
\begin{equation}
{\widehat D}_{n} b_{n}(\tau) + {\widehat E}_{n} a_{n}(\tau)
+ {\widehat F}_{n} d_{n}(\tau) + {\widehat G}_{n} e_{n}(\tau)
= 0 ,
\label{(9.2.6)}
\end{equation}
\begin{equation}
{\widehat L}_{n} d_{n}(\tau)
+ {\widehat M}_{n} b_{n}(\tau) = 0,
\label{(9.2.7)}
\end{equation}
\begin{equation}
{\widehat N}_{n} e_{n}(\tau) + {\widehat P}_{n} b_{n}(\tau)
+ {\widehat Q}_{n} a_{n}(\tau) = 0,
\label{(9.2.8)}
\end{equation}
\begin{equation}
{\widehat H}_{n} c_{n}(\tau)
+ {\widehat K}_{n} f_{n}(\tau) = 0,
\label{(9.2.9)}
\end{equation}
\begin{equation}
{\widehat R}_{n} f_{n}(\tau)
+ {\widehat S}_{n} c_{n}(\tau) = 0,
\label{(9.2.10)}
\end{equation}
\begin{equation}
{\widehat T}_{n} k_{n}(\tau) = 0.
\label{(9.2.11)}
\end{equation}
Since our background is flat, after setting
$\alpha=1$ in (\ref{(9.2.1)}) the operators appearing in
Eqs. (\ref{(9.2.5)})--(\ref{(9.2.11)}) take the
form (for all integer $n \geq 3$)
\begin{equation}
{\widehat A}_{n} \equiv
{d^{2}\over d \tau^{2}} + {3\over \tau} {d\over d\tau}
- {(n^{2} + 5)\over \tau^{2}} +  \lambda_{n},
\label{(9.2.12)}
\end{equation}
\begin{equation}
{\widehat B}_{n} \equiv {4\over \tau^{3}},
\label{(9.2.13)}
\end{equation}
\begin{equation}
{\widehat C}_{n} \equiv {2\over \tau^{4}},
\label{(9.2.14)}
\end{equation}
\begin{equation}
{\widehat D}_{n} \equiv
{d^{2} \over d\tau^{2}} + {1\over \tau} {d\over d\tau}
- {(n^{2} + 4)\over \tau^{2}} +  \lambda_{n},
\label{(9.2.15)}
\end{equation}
\begin{equation}
{\widehat E}_{n} \equiv {2\over \tau} (n^{2} - 1),
\label{(9.2.16)}
\end{equation}
\begin{equation}
{\widehat F}_{n} \equiv {4\over 3} {(n^{2} - 4)\over \tau^{3}},
\label{(9.2.17)}
\end{equation}
\begin{equation}
{\widehat G}_{n} \equiv - {2\over 3} {(n^{2} - 1)\over \tau^{3}},
\label{(9.2.18)}
\end{equation}
\begin{equation}
{\widehat H}_{n} \equiv {d^{2}\over d \tau^{2}}
+ {1\over \tau} {d\over d\tau}
- {(n^{2} + 5)\over \tau^{2}} + \lambda_{n},
\label{(9.2.19)}
\end{equation}
\begin{equation}
{\widehat K}_{n} \equiv {2\over \tau^{3}} (n^{2} - 4),
\label{(9.2.20)}
\end{equation}
\begin{equation}
{\widehat L}_{n} \equiv {d^{2}\over d\tau^{2}}
- {1\over \tau} {d\over d\tau}
- {(n^{2} - 5)\over \tau^{2}} +  \lambda_{n},
\label{(9.2.21)}
\end{equation}
\begin{equation}
{\widehat M}_{n} \equiv {4\over \tau},
\label{(9.2.22)}
\end{equation}
\begin{equation}
{\widehat N}_{n} \equiv {d^{2}\over d \tau^{2}}
- {1\over \tau} {d\over d\tau}
- {(n^{2} + 1)\over \tau^{2}} +  \lambda_{n},
\label{(9.2.23)}
\end{equation}
\begin{equation}
{\widehat P}_{n} \equiv - {4\over \tau},
\label{(9.2.24)}
\end{equation}
\begin{equation}
{\widehat Q}_{n} \equiv 6,
\label{(9.2.25)}
\end{equation}
\begin{equation}
{\widehat R}_{n} \equiv {d^{2}\over d\tau^{2}}
- {1\over \tau} {d\over d\tau}
- {(n^{2} - 4)\over \tau^{2}} +  \lambda_{n},
\label{(9.2.26)}
\end{equation}
\begin{equation}
{\widehat S}_{n} \equiv {2\over \tau},
\label{(9.2.27)}
\end{equation}
\begin{equation}
{\widehat T}_{n} \equiv {d^{2}\over d\tau^{2}}
- {1\over \tau} {d\over d\tau}
- {(n^{2} - 1)\over \tau^{2}} +  \lambda_{n}.
\label{(9.2.28)}
\end{equation}
Inserting the operator ${\widehat T}_{n}$
from Eq. (\ref{(9.2.28)}) into Eq. (\ref{(9.2.11)})
one can easily find the basis function describing the
transverse-traceless symmetric tensor harmonics which usually are
treated as physical degrees of freedom:
\begin{equation}
k_{n}(\tau) = \alpha_{1} \tau I_{n}(M \tau)
+ \alpha_{2} \tau K_{n}(M \tau) \; , \;  n=3,\ldots ,
\label{(9.2.29)}
\end{equation}
where $M = \sqrt{-\lambda}$ and $I$ and $K$
are modified Bessel functions.

However, the equations (\ref{(9.2.5)})--(\ref{(9.2.8)})
for scalar-type gravitational
perturbations lead to a rather complicated entangled system as well
as Eqs. (\ref{(9.2.9)}) and (\ref{(9.2.10)}),
describing vector perturbations. In
Refs. \cite{EKMP1994,EKMP95a},
where the analogous problem was studied for the
electromagnetic field, a method was used to decouple a similar
entangled system for normal and longitudinal components of
the 4-vector potential. The idea
is that one can diagonalize a $2 \times 2$ operator matrix after
multiplying it by two functional matrices. In some cases
one can choose these functional matrices in such a way that the
transformed operator matrix is diagonal and the corresponding
differential equations for basis functions are decoupled.
However, in the case of scalar-type gravitational perturbations we
have a $4 \times 4$ operator matrix.
To diagonalize such a matrix it is
necessary to solve a system of 24 second-order algebraic
equations with 24 variables. This problem seems a rather cumbersome
one and we thus use another method.
For this purpose, we assume that the solution of
the system of equations (\ref{(9.2.5)})--(\ref{(9.2.8)})
is some set of modified Bessel
functions with unknown index $\nu$. Let us look for a solution of this
system in the form
\begin{equation}
a_{n}(\tau) = \beta_{1} {W_{\nu}(M\tau)\over \tau},
\label{(9.2.30)}
\end{equation}
\begin{equation}
b_{n}(\tau) = \beta_{2} W_{\nu}(M\tau),
\label{(9.2.31)}
\end{equation}
\begin{equation}
d_{n}(\tau) = \beta_{3} \tau W_{\nu}(M\tau),
\label{(9.2.32)}
\end{equation}
\begin{equation}
e_{n}(\tau) = \beta_{4} \tau W_{\nu}(M\tau).
\label{(9.2.33)}
\end{equation}
Here, $W_{\nu}$ is a linear combination of modified Bessel functions
$I_{\nu}$ and $K_{\nu}$ obeying the Bessel equation
\begin{equation}
\left({d^{2}\over d\tau^{2}} + {1\over \tau}{d\over d\tau}
- {\nu^{2}\over \tau^{2}} -  M^{2}\right) W_{\nu}(M\tau) = 0.
\label{(9.2.34)}
\end{equation}
Now, inserting the functions (\ref{(9.2.30)})--(\ref{(9.2.33)})
and the corresponding operators into the system of equations
(\ref{(9.2.5)})--(\ref{(9.2.8)}), and taking into
account the Bessel equation (\ref{(9.2.34)}),
one finds the following system of equations for $\beta_{1},
\beta_{2}, \beta_{3}$ and $\beta_{4}$,
\begin{equation}
(\nu^{2} - n^{2} - 6)\beta_{1} + 4\beta_{2} + 2\beta_{4} = 0,
\label{(9.2.35a)}
\end{equation}
\begin{equation}
6(n^{2} - 1)\beta_{1} +3(\nu^{2} - n^{2} - 4)\beta_{2}
+ 4(n^{2} - 4)\beta_{3} - 2(n^{2} - 1)\beta_{4} =0,
\label{(9.2.35b)}
\end{equation}
\begin{equation}
4\beta_{2} + (\nu^{2} - n^{2} +4)\beta_{3} = 0,
\label{(9.2.35c)}
\end{equation}
\begin{equation}
6\beta_{1} - 4\beta_{2} + (\nu^{2} - n^{2} -2)\beta_{4} = 0.
\label{(9.2.35d)}
\end{equation}
The condition for the existence of non-trivial solutions of
the linear homogeneous system
(\ref{(9.2.35a)})--(\ref{(9.2.35d)})
is the vanishing of its determinant, i.e.
\begin{equation}
(\nu^{2} - n^{2})^{2}
\Bigr[(\nu^{2} - n^{2})^{2} - 8(\nu^{2} - n^{2})
- 16(n^{2} - 1)\Bigr] = 0.
\label{(9.2.36)}
\end{equation}
The roots of Eq. (\ref{(9.2.36)}) are
$$
\nu^{2} = n^{2},\ \nu^{2} = (n-2)^{2},\ \nu^{2} = (n+2)^{2}.
$$
The positive values of $\nu$ provide the orders of modified
Bessel functions.
Now we can write down the $\beta$'s corresponding to
different values for $\nu$'s. For $\nu = n$ one has
\begin{equation}
\beta_{4} = 3\beta_{1},\ \beta_{2} = \beta_{3} = 0,
\label{(9.2.37)}
\end{equation}
or
\begin{equation}
\beta_{1} = 0,\ \beta_{3} = -\beta_{2},
\beta_{4} = -2\beta_{2}.
\label{(9.2.38)}
\end{equation}
For $\nu = n - 2$ one has
\begin{equation}
\beta_{2} = (n + 1)\beta_{1},\ \beta_{3} = {(n + 1)\over (n - 2)}
\beta_{1},\ \beta_{4} = -\beta_{1}.
\label{(9.2.39)}
\end{equation}
Last, for  $\nu = n + 2$ one has
\begin{equation}
\beta_{2} = -(n - 1)\beta_{1},\ \beta_{3} = {(n - 1)\over (n + 2)}
\beta_{1},\ \beta_{4} = -\beta_{1}.
\label{(9.2.40)}
\end{equation}
Having the Eqs. (\ref{(9.2.37)})--(\ref{(9.2.40)})
we can get the basis functions for
scalar-type gravitational perturbations
(\ref{(9.2.30)})--(\ref{(9.2.33)}),
\begin{eqnarray}
a_{n}(\tau)&=& {1\over \tau}\Bigr[\gamma_{1}I_{n}(M\tau)
+\gamma_{3}I_{n-2}(M\tau) + \gamma_{4}I_{n+2}(M\tau) \nonumber \\
&+& \delta_{1}K_{n}(M\tau) +\delta_{3}K_{n-2}(M\tau) +
\delta_{4}K_{n+2}(M\tau) \Bigr],
\label{(9.2.41)}
\end{eqnarray}
\begin{eqnarray}
b_{n}(\tau)&=& \gamma_{2}I_{n}(M\tau) + (n + 1)
\gamma_{3}I_{n-2}(M\tau) \nonumber \\
&-& (n-1)\gamma_{4}I_{n+2}(M\tau)
+\delta_{2}K_{n}(M\tau) \nonumber \\
&+& (n+1)\delta_{3}K_{n-2}(M\tau)
- (n-1)\delta_{4}K_{n+2}(M\tau),
\label{(9.2.42)}
\end{eqnarray}
\begin{eqnarray}
d_{n}(\tau)&=& \tau \left[-\gamma_{2}I_{n}(M\tau) +
{(n+1)\over (n-2)}\gamma_{3}I_{n-2}(M\tau) \right. \nonumber \\
&+& {(n-1)\over (n+2)}\gamma_{4}I_{n+2}(M\tau)
-\delta_{2}K_{n}(M\tau) \nonumber \\
&\;& \left. +{(n+1)\over (n-2)}\delta_{3}K_{n-2}(M\tau)
+ {(n-1)\over (n+2)}\delta_{4}K_{n+2}(M\tau)\right],
\label{(9.2.43)}
\end{eqnarray}
\begin{eqnarray}
e_{n}(\tau)&=&\tau
\Bigr[3\gamma_{1}I_{n}(M\tau) - 2\gamma_{2}I_{n}(M\tau)
-\gamma_{3}I_{n-2}(M\tau) \nonumber \\
&-& \gamma_{4}I_{n+2}(M\tau)
+3\delta_{1}K_{n}(M\tau) - 2\delta_{2}K_{n}(M\tau) \nonumber \\
&-& \delta_{3}K_{n-2}(M\tau)
- \delta_{4}K_{n+2}(M\tau)\Bigr].
\label{(9.2.44)}
\end{eqnarray}

We can find the basis functions for vectorlike gravitational
perturbations in a similar way. Let us suppose that
\begin{equation}
c_{n}(\tau) = \varepsilon_{1} W_{\nu}(M\tau),
\label{(9.2.45)}
\end{equation}
and
\begin{equation}
f_{n}(\tau) = \varepsilon_{2} \tau W_{\nu}(M\tau).
\label{(9.2.46)}
\end{equation}
Inserting (\ref{(9.2.45)}) and (\ref{(9.2.46)}) into Eqs.
(\ref{(9.2.9)}) and (\ref{(9.2.10)}) one has the system
$$
(\nu^{2} - n^{2} - 5)\varepsilon_{1}
+ 2(n^{2} - 4)\varepsilon_{2} = 0,
$$
\begin{equation}
2\varepsilon_{1} + (\nu^{2} - n^{2} + 3)\varepsilon_{2} = 0.
\label{(9.2.47)}
\end{equation}
The determinant of the system (\ref{(9.2.47)}) is
\begin{equation}
(\nu^{2} - n^{2})^{2} - 2(\nu^{2} - n^{2}) - 4n^{2} + 1,
\end{equation}
and its positive roots are $n \pm 1$.
For $\nu = n + 1 $ one has
$$
\varepsilon_{2} = -{1\over (n + 2)} \varepsilon_{1},
$$
and for $\nu = n - 1 $ one has
$$
\varepsilon_{2} = {1\over (n - 2)} \varepsilon_{1} ,
$$
and correspondingly the basis functions (\ref{(9.2.45)}) and
(\ref{(9.2.46)}) take the form
\begin{equation}
c_{n}(\tau) = {\widetilde \varepsilon}_{1}I_{n+1}(M\tau) +
{\widetilde \varepsilon}_{2}I_{n-1}(M\tau) +
\eta_{1}K_{n+1}(M\tau) +
\eta_{2}K_{n-1}(M\tau) ,
\label{(9.2.48)}
\end{equation}
\begin{eqnarray}
f_{n}(\tau)&=& \tau\left[-{1\over (n+2)}
{\widetilde \varepsilon}_{1}I_{n+1}(M\tau)
+ {1\over (n-2)}{\widetilde \varepsilon}_{2}
I_{n-1}(M\tau)\right. \nonumber \\
&\;& \left. - {1\over (n+2)}\eta_{1}K_{n+1}(M\tau) +
{1\over (n-2)}\eta_{2}K_{n-1}(M\tau)\right].
\label{(9.2.49)}
\end{eqnarray}

We have also to find the basis functions for ghosts. The eigenvalue
equations for ghosts in the de Donder gauge have the form
$$
\cstok{\ }\varphi_{\mu}^{(\lambda)}
+ \lambda \varphi_{\mu}^{(\lambda)} = 0,
$$
and the corresponding fields can be expanded on a family of
3-spheres as
\begin{equation}
\varphi_{0}(x,\tau) = \sum_{n=1}^{\infty} l_{n}(\tau) Q^{(n)}(x),
\label{(9.2.50)}
\end{equation}
\begin{equation}
\varphi_{i}(x,\tau) = \sum_{n=2}^{\infty} \biggr[
m_{n}(\tau) {Q_{\mid i}^{(n)}(x)\over (n^{2} - 1)}
+p_{n}(\tau)S_{i}^{(n)}(x)\biggr].
\label{(9.2.51)}
\end{equation}
The functions $l_{n}(\tau), m_{n}(\tau)$ and $p_{n}(\tau)$
can be found similarly to those for
harmonics of gravitational perturbations. They have the form
\begin{equation}
l_{n}(\tau) = {1\over \tau}\Bigr[\kappa_{1}I_{n+1}(M\tau) +
\kappa_{2}I_{n-1}(M\tau)
+ \theta_{1}K_{n+1}(M\tau) + \theta_{2}K_{n-1}(M\tau)\Bigr] ,
\label{(9.2.52)}
\end{equation}
\begin{eqnarray}
m_{n}(\tau)&=& - (n-1)\kappa_{1}I_{n+1}(M\tau) +
(n+1)\kappa_{2}I_{n-1}(M\tau) \nonumber \\
&-& (n-1)\theta_{1}K_{n+1}(M\tau) +
(n+1)\theta_{2}K_{n-1}(M\tau),
\label{(9.2.53)}
\end{eqnarray}
\begin{equation}
p_{n}(\tau) = \vartheta I_{n}(M\tau) + \rho K_{n}(M\tau).
\label{(9.2.54)}
\end{equation}

\section{Barvinsky boundary conditions}

As we know from section 5, one can set to zero
at the boundary the gauge-averaging functional, the whole
ghost 1-form, and the perturbation of the induced 3-metric.
With the notation of section 5, after making an infinitesimal
{\it gauge} transformation of the metric perturbation $h_{\mu \nu}$
according to the law (\ref{(6.5.1)}),
one finds in the de Donder gauge (cf. (\ref{(6.5.7)}))
\begin{equation}
\Phi_{\mu}^{dD}(h)-\Phi_{\mu}^{dD}(\widehat h)=-{1\over 2}
\Bigr(\delta_{\mu}^{\; \; \nu}\cstok{\ }
+R_{\mu}^{\; \; \nu}\Bigr)\varphi_{\nu},
\label{(9.5.2)}
\end{equation}
where $R_{\mu \nu}$ is the Ricci tensor of the background,
and the elliptic operator on the right-hand side of
(\ref{(9.5.2)}) acts linearly on the ghost 1-form.
In our flat Euclidean background, the Ricci tensor vanishes,
and on making a 3+1 split of the de Donder functional
$\Phi_{\nu}^{dD}$ and of the ghost 1-form
$\varphi_{\mu}$, the boundary conditions proposed in
\cite{B1987} read as (unlike section 5, we here consider only
the de Donder gauge-averaging functional, with the corresponding
superscript $dD$)
\begin{equation}
\Bigr[h_{ij}\Bigr]_{\partial M}
=\Bigr[{\widehat h}_{ij}\Bigr]_{\partial M}=0,
\label{(9.5.3)}
\end{equation}
\begin{equation}
\Bigr[\Phi_{0}^{dD}(h)\Bigr]_{\partial M}
=\Bigr[\Phi_{0}^{dD}(\widehat h)\Bigr]_{\partial M}=0,
\label{(9.5.4)}
\end{equation}
\begin{equation}
\Bigr[\Phi_{i}^{dD}(h)\Bigr]_{\partial M}
=\Bigr[\Phi_{i}^{dD}(\widehat h)\Bigr]_{\partial M}=0,
\label{(9.5.5)}
\end{equation}
\begin{equation}
\Bigr[\varphi_{0}\Bigr]_{\partial M}=0,
\label{(9.5.6)}
\end{equation}
\begin{equation}
\Bigr[\varphi_{i}\Bigr]_{\partial M}=0.
\label{(9.5.7)}
\end{equation}
Once again, the vanishing of the whole ghost 1-form at the
boundary ensures the invariance of the boundary
conditions (\ref{(9.5.3)}) under the transformations
(\ref{(6.5.1)}). At that
stage, the only remaining set of boundary conditions on
metric perturbations, whose invariance
under (\ref{(6.5.1)}) is again
guaranteed by (\ref{(9.5.6)}), (\ref{(9.5.7)}),
is given by (\ref{(9.5.4)}), (\ref{(9.5.5)})
by virtue of (\ref{(9.5.2)}).
In this respect, these boundary conditions are the
natural generalization of magnetic boundary conditions for
Euclidean Maxwell theory, where one sets to zero at the
boundary the tangential components of the potential, the
gauge-averaging functional, and hence the ghost 0-form.
The boundary conditions (\ref{(9.5.3)})--(\ref{(9.5.7)})
were considered in \cite{B1987} as part of
the effort to understand the relation
between the wave function of the universe and the effective
action in quantum field theory. The loop expansion in
quantum cosmology was then obtained after a thorough study
of boundary conditions for the propagator.

In light of (\ref{(9.5.3)}), the boundary
conditions (\ref{(9.5.4)}), (\ref{(9.5.5)})
lead to mixed boundary conditions on the metric perturbations
which take the form (cf. \cite{MAO1991,AE1998,AE1999})
\begin{equation}
\left[{\partial h_{00}\over \partial \tau}
+{6\over \tau}h_{00}-{\partial \over \partial \tau}
\Bigr(g^{ij}h_{ij}\Bigr)+{2\over \tau^{2}}
h_{0i}^{\; \; \; \mid i} \right]_{\partial M}=0,
\label{(9.5.8)}
\end{equation}
\begin{equation}
\left[{\partial h_{0i}\over \partial \tau}
+{3\over \tau}h_{0i}-{1\over 2}
{\partial h_{00}\over \partial x^{i}}\right]_{\partial M}
=0.
\label{(9.5.9)}
\end{equation}
To evaluate the scaling behaviour of the corresponding
1-loop amplitudes, it is necessary to write down the
mode-by-mode form of the boundary conditions
(\ref{(9.5.8)}), (\ref{(9.5.9)}), (\ref{(9.5.3)}),
(\ref{(9.5.6)}) and (\ref{(9.5.7)}). They lead to
\begin{equation}
{da_{n}\over d\tau}+{6\over \tau}a_{n}-{1\over \tau^{2}}
{de_{n}\over d\tau}-{2\over \tau^{2}}b_{n}=0
\; \; \; \; {\rm at} \; \; \; \; {\partial M},
\label{(9.5.10)}
\end{equation}
\begin{equation}
{db_{n}\over d\tau}+{3\over \tau}b_{n}
-{(n^{2}-1)\over 2}a_{n}=0 \; \; \; \; {\rm at} \; \; \; \;
{\partial M} ,
\label{(9.5.11)}
\end{equation}
\begin{equation}
{dc_{n}\over d\tau}+{3\over \tau}c_{n}=0
\; \; \; \; {\rm at} \; \; \; \; {\partial M},
\label{(9.5.12)}
\end{equation}
\begin{equation}
d_{n}=0 \; \; \; \; {\rm at} \; \; \; \; {\partial M},
\label{(9.5.13)}
\end{equation}
\begin{equation}
e_{n}=0 \; \; \; \; {\rm at} \; \; \; \; {\partial M},
\label{(9.5.14)}
\end{equation}
\begin{equation}
f_{n}=0 \; \; \; \; {\rm at} \; \; \; \; {\partial M},
\label{(9.5.15)}
\end{equation}
\begin{equation}
k_{n}=0 \; \; \; \; {\rm at} \; \; \; \; {\partial M},
\label{(9.5.16)}
\end{equation}
\begin{equation}
l_{n}=0 \; \; \; \; {\rm at} \; \; \; \; {\partial M},
\label{(9.5.17)}
\end{equation}
\begin{equation}
m_{n}=0 \; \; \; \; {\rm at} \; \; \; \; {\partial M},
\label{(9.5.18)}
\end{equation}
\begin{equation}
p_{n}=0 \; \; \; \; {\rm at} \; \; \; \; {\partial M}.
\label{(9.5.19)}
\end{equation}
On using, for example, the technique of section 2, the corresponding
contributions to $\zeta(0)$ are found to be
(the results quoted below are independent of the particular
algorithm used)
\begin{equation}
\zeta(0)_{\rm transverse-traceless \; modes}=-{278\over 45},
\label{(9.5.20)}
\end{equation}
\begin{equation}
\zeta(0)_{\rm partially \; decoupled \; modes}=-2-15=-17,
\label{(9.5.22)}
\end{equation}
\begin{equation}
\zeta(0)_{\rm vector \; modes}=12-{11\over 60}
-{2\over 3}-{31\over 180}={494\over 45},
\label{(9.5.23)}
\end{equation}
\begin{equation}
\zeta(0)_{\rm decoupled \; vector \; mode}=-{15\over 2},
\label{(9.5.24)}
\end{equation}
\begin{equation}
\zeta(0)_{\rm scalar \; ghost \; modes}=-2\biggr({179\over 120}
+{59\over 360}\biggr)=-{149\over 45},
\label{(9.5.25)}
\end{equation}
\begin{equation}
\zeta(0)_{\rm vector \; ghost \; modes}=-2\biggr(-{41\over 120}
-{31\over 360}\biggr)={77\over 90},
\label{(9.5.26)}
\end{equation}
\begin{equation}
\zeta(0)_{\rm decoupled \; ghost \; mode}={5\over 2}.
\label{(9.5.27)}
\end{equation}

\subsection{Eigenvalue condition for scalar modes}

For scalar modes, which are not discussed in the previous list
of results, one finds eventually the
eigenvalues $E=X^{2}$ from the roots $X$ of
\cite{EFKK05a,EFKK05b}
\begin{equation}
J_{n}'(x) \pm {n \over x}J_{n}(x)=0,
\label{(87)}
\end{equation}
\begin{equation}
J_{n}'(x)+\left(-{x \over 2} \pm {n \over x}\right)J_{n}(x)=0,
\label{(88)}
\end{equation}
where $J_{n}$ are the Bessel functions of first kind.
Note that both $x$ and $-x$ solve the same equation.

\subsection{Four spectral zeta-functions for scalar modes}

As we know from section 2,
by virtue of the Cauchy theorem and of suitable rotations of
integration contours in the complex plane \cite{BKK1992,BGKE1996},
the eigenvalue conditions (\ref{(87)}) and (\ref{(88)}) give
rise to the following four spectral zeta-functions
\cite{EFKK05a,EFKK05b},
\begin{equation}
\zeta_{A,B}^{\pm}(s) \equiv {\sin(\pi s) \over \pi}
\sum_{n=3}^{\infty}n^{-(2s-2)}
\int_{0}^{\infty}dz {{\partial \over \partial z}
\log F_{A,B}^{\pm}(zn) \over z^{2s}},
\label{(89)}
\end{equation}
where, denoting by $I_{n}$ the modified Bessel functions of
the first kind (here $\beta_{+} \equiv n, \beta_{-} \equiv n+2$),
\begin{equation}
F_{A}^{\pm}(zn) \equiv z^{-\beta_{\pm}}
\Bigr(zn I_{n}'(zn) \pm n I_{n}(zn)\Bigr),
\label{(90)}
\end{equation}
\begin{equation}
F_{B}^{\pm}(zn) \equiv z^{-\beta_{\pm}}\left(zn I_{n}'(zn)
+\left({z^{2}n^{2}\over 2} \pm n \right)I_{n}(zn)\right).
\label{(91)}
\end{equation}
Regularity at the origin is easily proved in the elliptic sectors,
corresponding to $\zeta_{A}^{\pm}(s)$ and $\zeta_{B}^{-}(s)$
\cite{EFKK05a,EFKK05b}.

\subsection{Regularity at the origin of $\zeta_{B}^{+}$}

With the notation in Refs. \cite{EFKK05a,EFKK05b}, if one defines
the variable $\tau \equiv (1+z^{2})^{-{1\over 2}}$, one can write
the uniform asymptotic expansion of $F_{B}^{+}$ in the form
\cite{EFKK05a,EFKK05b}
\begin{equation}
F_{B}^{+} \sim {{\rm e}^{n \eta(\tau)}\over h(n)\sqrt{\tau}}
{(1-\tau^{2})\over \tau} \left(1+\sum_{j=1}^{\infty}
{r_{j,+}(\tau)\over n^{j}}\right).
\label{(92)}
\end{equation}
On splitting the integral $\int_{0}^{1}d\tau=\int_{0}^{\mu}d\tau
+\int_{\mu}^{1}d\tau$ with $\mu$ small, one gets an asymptotic expansion
of the left-hand side of Eq. (\ref{(89)}) by writing, in the first
interval on the right-hand side,
\begin{equation}
\log \left(1+\sum_{j=1}^{\infty}{r_{j,+}(\tau)\over n^{j}}\right)
\sim \sum_{j=1}^{\infty}{R_{j,+}(\tau)\over n^{j}},
\label{(93)}
\end{equation}
and then computing \cite{EFKK05a,EFKK05b}
\begin{equation}
C_{j}(\tau) \equiv {\partial R_{j,+}\over \partial \tau}
=(1-\tau)^{-j-1}\sum_{a=j-1}^{4j}K_{a}^{(j)}\tau^{a}.
\label{(94)}
\end{equation}
Remarkably, by virtue of the identity obeyed by the spectral
coefficients $K_{a}^{(j)}$ {\it on the 4-ball}, i.e.
\begin{equation}
g(j) \equiv \sum_{a=j}^{4j}{\Gamma(a+1)\over \Gamma(a-j+1)}
K_{a}^{(j)}=0,
\label{(95)}
\end{equation}
which holds $\forall j=1,...,\infty$, one finds
\cite{EFKK05a,EFKK05b}
\begin{equation}
\lim_{s \to 0} s \zeta_{B}^{+}(s)={1\over 6}\sum_{a=3}^{12}
a(a-1)(a-2)K_{a}^{(3)}=0,
\label{(96)}
\end{equation}
and \cite{EFKK05a,EFKK05b}
\begin{equation}
\zeta_{B}^{+}(0)={5\over 4}+{1079 \over 240}-{1\over 2}
\sum_{a=2}^{12}\omega(a)K_{a}^{(3)}+\sum_{j=1}^{\infty}f(j)g(j)
={296\over 45},
\label{(97)}
\end{equation}
where, on denoting here by $\psi$ the logarithmic
derivative of the $\Gamma$-function \cite{EFKK05a,EFKK05b},
\begin{eqnarray}
\omega(a) & \equiv & {1\over 6}{\Gamma(a+1)\over \Gamma(a-2)}
\biggr[-\log(2)-{(6a^{2}-9a+1)\over 4}{\Gamma(a-2)\over \Gamma(a+1)}
\nonumber \\
&+& 2\psi(a+1)-\psi(a-2)-\psi(4)\biggr],
\label{(98)}\\
f(j) &\equiv & {(-1)^{j}\over j!}\Bigr[-1-2^{2-j}
+\zeta_{R}(j-2)(1-\delta_{j,3})+\gamma \delta_{j,3}\Bigr].
\label{(99)}
\end{eqnarray}
Equation (\ref{(95)}) achieves three goals:
\vskip 0.3cm
\noindent
(i) Vanishing of the $\log(2)$ coefficient in (\ref{(97)});
\vskip 0.3cm
\noindent
(ii) Vanishing of $\sum_{j=1}^{\infty}f(j)g(j)$ in (\ref{(97)});
\vskip 0.3cm
\noindent
(iii) Regularity at the origin of $\zeta_{B}^{+}$.

\subsection{Interpretation of the result}

Since all other $\zeta(0)$ values for pure gravity obtained in the
literature on the 4-ball are negative,
the analysis here briefly outlined shows that only fully
diffeomorphism-invariant boundary conditions lead to a positive
$\zeta(0)$ value for pure gravity on the 4-ball, and hence {\it only
fully diffeomorphism-invariant boundary conditions lead to a
vanishing cosmological wave function for vanishing 3-geometries
at 1-loop level, at least on the Euclidean 4-ball}. If the
probabilistic interpretation is tenable for the whole universe,
this means that {\it the universe has vanishing probability of reaching
the initial singularity} at $a=0$,
which is therefore avoided by virtue
of quantum effects \cite{EFKK05a,EFKK05b}, since the 1-loop
wave function is proportional to $a^{\zeta(0)}$ \cite{S1985}.

\section{The strong ellipticity issue}

The result outlined in section 7 is non-trivial because the
zeta-function $\zeta_{B}^{+}$ corresponds to the sector of the
boundary-value problem for which strong ellipticity
\cite{K2001} fails to hold \cite{AE1998,AE1999}.
We now define in detail this concept,
and we are also going to discuss its relevance both for physics
and mathematics.

Let $M$  be a smooth, compact Riemannian manifold endowed with a
positive-definite metric $g$ and assume that the boundary $\partial M$
is smooth. Let $V$ be a vector bundle over $M$ and $C^{\infty}(V,M)$
be the space of smooth sections of the bundle $V$. With the introduction
of a Hermitian metric $E$ and the Riemannian volume element on $M$, the
dual bundle $V^{\ast}$ can be identified with $V$ and a natural inner
product for the smooth sections of $V$ can be defined. It is clear that
the Hilbert space $\mathcal{L}^{2}(V,M)$ is identified with the completion
of $C^{\infty}(V,M)$ with respect to the inner product. An operator of
Laplace type is a map
\begin{equation}
\mathscr{L}: C^{\infty}(V,M)\longrightarrow C^{\infty}(V,M)\;,
\end{equation}
expressed as follows:
\begin{equation}\label{wil1}
\mathscr{L}=-g^{ab}\nabla_{a}^{V}\nabla_{b}^{V}+Q\;,
\end{equation}
where $\nabla^{V}$ denotes the connection on $V$ and $Q$ represents
a self-adjoint endomorphism of $V$.

The boundary data for the Laplace operator under consideration can be
written as
\begin{equation}
\psi_{\mathscr{L}}(\varphi)=\left(
\begin{array}{c}
\psi_{0}(\varphi) \\
\psi_{1}(\varphi) \\
\end{array}
\right)\;,
\end{equation}
where $\varphi\in C^{\infty}(V,M)$ and we have set
\begin{equation}
\psi_{0}(\varphi)=\varphi|_{\partial M}\;,\quad \psi_{1}
(\varphi)=\nabla_{N}\varphi|_{\partial M}\;,
\end{equation}
with $\nabla_{N}$ denoting the normal covariant derivative with respect
to the boundary $\partial M$. By introducing the tangential differential
operator $B_{\mathscr{L}}$ on $\partial M$ the boundary conditions can
be written in a concise way as
\begin{equation}\label{wil2}
B_{\mathscr{L}}\psi_{\mathcal{L}}(\varphi)=0\;,
\end{equation}
where the general form of $B_{\mathscr{L}}$, which ensures the
self-adjointness of the operator $\mathcal{L}$,
is \cite{MAO1991,AE1998,AE1999}
\begin{equation}
B_{\mathscr{L}}=\left(
\begin{array}{cc}
\Pi & 0 \\
\Lambda & (\mathbb{I}-\Pi) \\
\end{array}
\right)\;,
\end{equation}
where $\Pi$ denotes a self-adjoint projector and $\Lambda$ is a
self-adjoint tangential differential
operator satisfying the relation
\begin{equation}
  \Pi\Lambda=\Lambda\Pi=0\;,
\end{equation}
and $\mathbb{I}$ is the identity endomorphism of $V$.
The operator $\Lambda$ can always be cast in the manifestly
self-adjoint form
\begin{equation}\label{wil3}
  \Lambda=(\mathbb{I}-\Pi)\left\{\frac{1}{2}\left(\Gamma^{i}
\hat{\nabla}_{i}+\hat{\nabla}_{i}\Gamma^{i}\right)
+S\right\}(\mathbb{I}-\Pi)\;,
\end{equation}
where $\hat{\nabla}_{i}$ represents the covariant tangential derivative,
compatible with the induced metric on $\partial M$,
and $\Gamma^{i}$ and $S$
are endomorphisms satisfying the relations
\begin{equation}\label{wil4}
\bar{\Gamma}^{i}=E^{-1}(\Gamma^{i})^\dagger E=-\Gamma^{i} \;,\quad
\bar{S}=E^{-1}S^\dagger E=S\;,
\end{equation}
\begin{equation}
\Pi\Gamma^{i}=\Gamma^{i}\Pi=\Pi S=S\Pi=0\;.
\end{equation}
Here, the bar denotes the adjoint in the space ${\mathcal L}^2 (V,M)$ 
and dagger is the Hermitian conjugate.

It is instructive to notice that different choices for the projector
$\Pi$ and the operator $\Lambda$ lead to
different types of boundary conditions.
More precisely: for $\Pi=\mathbb{I}$ and $\Lambda=0$ one obtains
Dirichlet boundary conditions, for $\Pi=\Lambda=0$
one recovers Neumann boundary conditions, and for
$\Pi=0$ and $\Lambda\neq 0$, with $\Lambda$ not a differential operator,
the boundary conditions are reduced to the Robin type.

The leading symbol of the Laplace operator (\ref{wil1}) is defined as
\begin{equation}\label{wil5}
\sigma_{L}(\mathscr{L};x,\xi) \equiv |\xi|^{2}\cdot\mathbb{I}
=g^{ab}(x)\xi_{a}\xi_{b}\cdot\mathbb{I}\;,
\end{equation}
where $\xi\in T^{\ast}M$ is an arbitrary cotangent vector. For a
non-singular Riemannian metric the leading symbol
(\ref{wil3}) is positive-definite and, hence,
the operator $\mathscr{L}$ is elliptic \cite{gilkey95}. In order to
analyze the strong ellipticity condition for the
boundary-value problem (\ref{wil1}) and
(\ref{wil2}) we need to introduce the leading symbol of the boundary-value
operator $B_{\mathscr{L}}$. Such leading symbol, denoted by
$\sigma_{g}(B_{\mathscr{L}})$,
is defined as \cite{AE1999}
\begin{equation}
\sigma_{g}(B_{\mathscr{L}})=\left(
\begin{array}{cc}
\Pi & 0 \\
iT & (\mathbb{I}-\Pi) \\
\end{array}
\right)\;,
\end{equation}
where, by exploiting the relation (\ref{wil3}), one has
\begin{equation}\label{wil5a}
T=-i\sigma_{L}(\Lambda)=\Gamma^{i}\rho_{i}\;,
\end{equation}
with an arbitrary cotangent vector $\rho$ on $T^{\ast}(\partial M)$,
the cotangent bundle of the boundary of $M$.
From the relations (\ref{wil4}) one can prove that the matrix $T$
defined above is anti-self-adjoint,
\begin{equation}
\bar{T}=-T\;,
\end{equation}
and it satisfies the equation
\begin{equation}
\Pi T=T\Pi=0\;.
\end{equation}

In a neighbourhood of $\partial M$ the Riemannian manifold $M$
can be locally described by a direct product $\Omega=[0,\epsilon]
\times\partial M$. A local set of coordinates for
$\Omega$ is $x^{\mu}=x^{\mu}(r,\hat{x}^{i})$
where $r$ denotes the normal distance from the boundary and $\hat{x}^{i}$
are the coordinates on the {\it moved}
boundary $\partial M(r)=\{x\in M\,| r(x)=r\}$ with $r\in [0,\epsilon]$.
In this setting, the leading symbol of the Laplace operator $\mathscr{L}$
can be written as $\sigma_{L}(\mathscr{L}; \hat{x},r,\rho,\omega)$.
Let us set $r=0$, make the replacement $\omega\to -i\partial_{r}$,
and consider the resulting differential equation
\begin{equation}\label{wil6}
\left[\sigma_{L}(\mathscr{L}; \hat{x},0,\rho,-i\partial_{r})
-\lambda\cdot\mathbb{I}\right]\phi(r)=0\;,
\end{equation}
with $\lambda\in\mathbb{C}-\mathbb{R}^{+}$, and whose
solutions must satisfy the asymptotic condition
\begin{equation}\label{wil7}
  \lim_{r\to\infty}\phi(r)=0\;.
\end{equation}

The boundary-value problem, consisting of the pair (\ref{wil1})
and (\ref{wil2}), is said to be {\it strongly elliptic}
with respect to the cone $\mathbb{C}-\mathbb{R}^{+}$ if, for
any $\rho\in T^{\ast}(\partial M)$, $\lambda\in\mathbb{C}
-\mathbb{R}^{+}$ and $\psi'_{\mathscr{L}}$,
there exists a unique solution to the equation (\ref{wil6}) satisfying
both the asymptotic condition (\ref{wil7}) and the relation
\begin{equation}\label{wil7a}
  \sigma_{g}(B_{\mathscr{L}})(\hat{x},\rho)\psi_{\mathscr{L}}(\phi)
=\psi_{\mathscr{L}}'(\phi)\;.
\end{equation}

For an operator of Laplace type the differential equation (\ref{wil6})
can be written as
\begin{equation}
\left[-\partial_{r}^{2}+|\rho|^{2}-\lambda\right]\phi(r)=0\;.
\end{equation}
The general solution to the above equation which satisfies the
condition (\ref{wil7}) of decay as $r\to\infty$ is
\begin{equation}\label{wil8}
  \phi(r)=\chi \exp(-\mu r)\;,
\end{equation}
where we have set $\mu=\sqrt{|\rho|^{2}-\lambda}$. The boundary data
for the solution (\ref{wil8})
can be expressed as
\begin{equation}\label{wil9}
\psi_{\mathscr{L}}(\phi)=\left(
\begin{array}{c}
\chi \\
-\mu\chi \\
\end{array}
\right)\;.
\end{equation}

The boundary-value problem under consideration is strongly elliptic
if the boundary data (\ref{wil9})
satisfy equation (\ref{wil7a}). More precisely, strong ellipticity
holds if the matrix associated with the linear system
\begin{equation}\label{wil9a}
\left(
\begin{array}{cc}
\Pi & 0 \\
iT & (\mathbb{I}-T) \\
\end{array}
\right)\left(
\begin{array}{c}
\chi \\
-\mu\chi \\
\end{array}
\right)=\left(
\begin{array}{c}
\psi'_{0} \\
\psi'_{1} \\
\end{array}
\right)
\end{equation}
is invertible. One can show that the above system is equivalent to the set
\begin{eqnarray}\label{wil10}
\Pi\chi&=&\psi'_{0},\nonumber\\
(\mu\mathbb{I}-iT)\chi&=&\mu\psi'_{0}-\psi'_{1}\;.
\end{eqnarray}
Since the first equation in (\ref{wil10}) is independent of the second
\cite{AE1999}, verifying the invertibility of the matrix in (\ref{wil9a}),
and hence strong ellipticity,
is equivalent to verifying the existence of a unique solution to the second
equation in (\ref{wil10}) for arbitrary $\psi'_{0}$ and $\psi'_{1}$.
A necessary and sufficient condition for the existence of a unique solution
to the second equation in (\ref{wil10}) can be found to be
\begin{equation}\label{wil11}
\det\left[\mu\mathbb{I}-iT\right]\neq 0\;.
\end{equation}

The matrix $iT$ defined in (\ref{wil5a}) is self-adjoint, and therefore
its eigenvalues are real. It is clear that for any $\lambda\in\mathbb{C}
-\mathbb{R}_{+}$
the quantity $\mu=\sqrt{|\rho|^{2}-\lambda}$ is complex and, hence,
the matrix $\left[\mu\mathbb{I}-iT\right]$ is non-degenerate. For
$\lambda\in\mathbb{R}_{-}$ the quantity
$\mu$ is real and satisfies the inequality $\mu>|\rho|$. The above remarks
together with the condition (\ref{wil11}) imply that
\begin{equation}
  |\rho|\mathbb{I}-iT>0\;.
\end{equation}
By noticing that $(|\rho|\mathbb{I}-iT)(|\rho|\mathbb{I}+iT)
=\mathbb{I}|\rho|^{2}+T^{2}$ we can conclude that the boundary-value
problem (\ref{wil1}) and (\ref{wil2})
is {\it strongly elliptic} if the eigenvalues of the matrix $T^{2}$ are
real and greater than $-|\rho|^{2}$, i.e.
\begin{equation}\label{wil12}
\Im(T^{2})=0\;,\quad \Re(\mathbb{I}|\rho|^{2}+T^{2})>0\;,
\end{equation}
for any $\rho\in T^{\ast}(\partial M)$.

In the setting of Euclidean quantum gravity it has been shown
\cite{AE1998,AE1999} that the eigenvalues of the matrix $T$ are the
following:
\begin{equation}
\textrm{spec}(T)=\left\{\begin{array}{ll}
0 &\quad \textrm{with degeneracy}\quad \left[m(m+1)/2-2\right]\\
i\rho &\quad \textrm{with degeneracy}\quad 1\\
-i\rho &\quad \textrm{with degeneracy}\quad 1\;.
\end{array}\right.
\end{equation}
Since the eigenvalues of the matrix $T^{2}$ are $0$ and $-|\rho|^{2}$
the strong ellipticity condition (\ref{wil12})
in {\it not} satisfied. This means, in particular,
that the dynamical operator of the metric perturbations
endowed with diffeomorphism-invariant boundary conditions in the de Donder
gauge is not strongly elliptic.

\section{Concluding remarks and open problems}

In order to avoid the problem of the lack of strong ellipticity in
Euclidean quantum gravity one can consider various alternative approaches.

When deriving the operator that describes the dynamics of the metric
perturbations the particular choice of the
de Donder gauge renders the operator of Laplace type but leads to the lack
of strong ellipticity of the associated boundary-value
problem. It remains to be seen whether strong ellipticity 
can be preserved if one considers instead
dynamical operators on metric perturbations which are non-minimal.

Another approach is to study Euclidean quantum gravity with non-local
boundary conditions \cite{esposito99}, or with
boundary conditions which are not gauge invariant and that eliminate the
occurrence of tangential derivatives \cite{luckock91}.

Although viable, the alternative approaches mentioned above contain some
problems and difficulties. It is, therefore, still unclear
what is the most appropriate way to solve the problem of the lack of strong
ellipticity in Euclidean quantum gravity. The result of section 7,
however, shows that there exists at least one background where a
meaningful $\zeta(0)$ value is still obtainable despite the lack of
strong ellipticity. At a deeper mathematical level, strong ellipticity
makes it possible to define the heat operator,
which is however not necessary
in order to define the resolvent or complex powers of the given
elliptic operator. For the latter two, one needs just a sector of the
complex plane free of eigenvalues of the leading symbol. [We are
grateful to Gerd Grubb for correspondence about this issue]. 
The investigation of other backgrounds 
might provide further examples of
meaningful $\zeta(0)$ values despite violation of strong ellipticity,
and their relevance for quantum cosmology and/or quantum field theory
should be assessed.\\[.3cm]

{\bf Acknowledgments} GE is indebted to Ivan Avramidi for scientific 
collaboration on strong ellipticity in quantum field theory and 
quantum gravity, and is grateful to the Dipartimento di Scienze Fisiche
of Federico II University, Naples, for hospitality and support. 
KK would like to thank Stuart Dowker for the many 
years of fruitful and very enjoyable collaboration.
Several of the papers written together and with others would never have 
seen the light of the day without his extremely valuable advice. 
The work of AK was partially supported by the RFBR grant
No 11-02-00643. KK is supported by the National Science Foundation 
Grant PHY-0757791.

\bibliographystyle{abbrv}

\begin{thebibliography}{100}
\bibitem{M1957}
Misner C W 1957 {\it Rev. Mod. Phys.} {\bf 29} 497
\bibitem{HH1983}
Hartle J B and Hawking S W 1983 {\it Phys. Rev.} D {\bf 28} 2960
\bibitem{H1984}
Hawking S W 1984 {\it Nucl. Phys.} B {\bf 239} 257
\bibitem{DC1976}
Dowker J S and Critchley R 1976 {\it Phys. Rev.} D {\bf 13} 3224
\bibitem{H1977}
Hawking S W 1977 {\it Commun. Math. Phys.} {\bf 55} 133
\bibitem{HT1992}
Henneaux M and Teitelboim C 1992 {\it Quantization of Gauge Systems}
(Princeton: Princeton University Press)
\bibitem{S1985}
Schleich K 1985 {\it Phys. Rev.} D {\bf 32} 1889
\bibitem{BKK1992}
Barvinsky A O, Kamenshchik A Yu and Karmazin I P 1992
{\it Ann. Phys. (N.Y.)} {\bf 219} 201
\bibitem{BKKM1992}
Barvinsky A O, Kamenshchik A Yu, Karmazin I P and Mishakov I V 1992
{\it Class. Quantum Grav.} {\bf 9} L27
\bibitem{KM1992}
Kamenshchik A Yu and Mishakov I V 1992
{\it Int J. Mod. Phys.} A {\bf 7} 3713
\bibitem{MP1990}
Moss I G and Poletti S 1990 {\it Nucl. Phys.} B {\bf 341} 155
\bibitem{EKMP1994}
Esposito G, Kamenshchik A Yu, Mishakov I V and Pollifrone G 1994
{\it Class. Quantum Grav.} {\bf 11} 2939
\bibitem{EKMP95a}
Esposito G, Kamenshchik A Yu, Mishakov I V and Pollifrone G 1995
{\it Phys. Rev.} D {\bf 52} 2183
\bibitem{BG1990}
Branson T P and Gilkey P B 1990 {\it Commun. Part. Diff. Eq.}
{\bf 15} 245
\bibitem{V1995}
Vassilevich D V 1995 {\it J. Math. Phys.} {\bf 36} 3174
\bibitem{MP1994}
Moss I G and Poletti S 1994 {\it Phys. Lett.} B {\bf 333} 326
\bibitem{EKMP95b}
Esposito G, Kamenshchik A Yu, Mishakov I V and Pollifrone G 1995
{\it Phys. Rev.} D {\bf 52} 3457
\bibitem{EFKK05a}
Esposito G, Fucci G, Kamenshchik A Yu and Kirsten K 2005
{\it Class. Quantum Grav.} {\bf 22} 957
\bibitem{EFKK05b}
Esposito G, Fucci, Kamenshchik A Yu and Kirsten K 2005
{\it J. High Energy Phys.} JHEP09(2005)063
\bibitem{bord96-37-895}
Bordag M, Elizalde E and Kirsten K 1996 {\it J. Math. Phys.} {\bf 37} 895
\bibitem{BGKE1996}
Bordag M, Geyer B, Kirsten K and Elizalde E 1996
{\it Commun. Math. Phys.} {\bf 179} 215
\bibitem{dowk96-13-585}
Dowker J S 1996 {\it Class. Quantum Grav.} {\bf 13} 585
\bibitem{dowk96-366-89}
Dowker J S 1996 {\it Phys. Lett.} B {\bf 366} 89
\bibitem{Olver}
Olver F W J 1974 {\it Introduction to Asymptotics and
Special Functions} (New York and London: Academic)
\bibitem{Thorne}
Thorne R C 1957 {\it Philos. Trans. R. Soc. London} {\bf 249} 597
\bibitem{K2001}
Kirsten K 2001 {\it Spectral Functions in Mathematics and Physics}
(Boca Raton: CRC Press)
\bibitem{eliz95b}
Elizalde E 1995 {\it Ten Physical Applications of Spectral Zeta Functions, 
Lecture Notes in Physics m35} (Berlin: Springer-Verlag)
\bibitem{bord09b}
Bordag M, Klimchitskaya GL, Mohideen U and Mostepanenko VM 2009 
{\it Advances in the Casimir Effect} (Oxford: Oxford University Press)
Kellogg O D 1954 {\it Foundations of Potential Theory}
(New York: Dover)
\bibitem{K1954}
Kellogg O D 1954 {\it Foundations of Potential Theory} (New York: Dover)
\bibitem{K1978}
Kennedy G 1978 {\it J. Phys.} A {\bf 11} L173
\bibitem{M1989}
Moss I G 1989 {\it Class. Quantum Grav.} {\bf 6} 759
\bibitem{MD1989}
Moss I G and Dowker J S 1989 {\it Phys. Lett.} B {\bf 229} 261
\bibitem{bord96-182-371}
Bordag M, Kirsten K and Dowker J S 1996 {\it Commun. Math. Phys.} 
{\bf 182} 371
\bibitem{abra70b}
Abramowitz M and Stegun I A 1970 {\it Handbook of Mathematical Functions}
(New York: Dover)
\bibitem{EKP1997}
Esposito G, Kamenshchik A Yu and Pollifrone G 1997
{\it Euclidean Quantum Gravity on Manifolds with Boundary},
{\it Fundamental Theories of Physics} {\bf 85} (Kluwer: Dordrecht)
\bibitem{B1987}
Barvinsky A O 1987 {\it Phys. Lett.} B {\bf 195} 344
\bibitem{DW2003}
DeWitt B S 2003 {\it The Global Approach to Quantum Field Theory},
International Series of Monographs on Physics {\bf 114}
(Oxford: Clarendon Press)
\bibitem{MAO1991}
McAvity D M and Osborn H 1991 {\it Class. Quantum Grav.}
{\bf 8} 1445
\bibitem{AE1998}
Avramidi I G and Esposito G 1998 {\it Class. Quantum Grav.}
{\bf 15} 1141
\bibitem{AE1999}
Avramidi I G and Esposito G 1999 {\it Commun. Math. Phys.}
{\bf 200} 495
\bibitem{gilkey95}
Gilkey P B 1995 {\it Invariance Theory, the Heat Equation and the
Atiyah--Singer Index Theorem} (Boca Raton: CRC Press)
\bibitem{esposito99}
Esposito G 1999 {\it Class. Quantum Grav.} {\bf 16} 1113
\bibitem{luckock91}
Luckock H C 1991 {\it J. Math. Phys.} {\bf 32} 1755
\end{thebibliography}

\end{document}